\documentclass[]{article}
\usepackage{amsfonts}
\usepackage{bm}
\usepackage[margin=1in]{geometry}
\usepackage[affil-it]{authblk}
\usepackage{graphicx}
\usepackage{bmpsize}
\usepackage{epstopdf}
\usepackage[square, numbers, comma, sort]{natbib}
\usepackage{amsmath}
\usepackage{xcolor}
\usepackage[toc,page]{appendix}
\usepackage{booktabs}
\usepackage{hyperref}

\renewcommand\appendix{\par
	\setcounter{section}{0}%
	\setcounter{subsection}{0}%
	\setcounter{Table}{0}
	\setcounter{figure}{0}
	\gdef\theTable{\Alph{Table}}
	\gdef\thefigure{\Alph{figure}}
	\section*{Appendix}
	\gdef\thesection{\Alph{section}}
	\setcounter{section}{1}}
\begin{document}
\title{
Edge waves and localisation in lattices containing tilted resonators}
\author[1,2]{Domenico Tallarico%
	\footnote{Corresponding author. Office 405,	Department of Mathematical Sciences, The University of Liverpool, L69 7ZL, Liverpool, United Kingdom. E-mail: \texttt{domenico.tallarico@liverpool.ac.uk}}}
\author[2]{Alessio Trevisan}
\author[1]{Natalia V. Movchan}
\author[1]{Alexander B. Movchan}
\affil[1]{Department of Mathematical 
	Sciences, Peach St, University of Liverpool, Liverpool L697ZL, United Kingdom.}
\affil[2]{EnginSoft SPA,
	Via Giambellino 7,
	 Padova 35129, Italy.}
\date{}
\maketitle
\begin{abstract}

	The paper presents the study of waves in a structured geometrically chiral solid. A special attention is given to the analysis of the Bloch-Floquet waves in a doubly periodic high-contrast lattice containing tilted resonators.  Dirac-like dispersion of Bloch waves in the structure is identified, studied and applied to wave-guiding and wave-defect interaction problems. The work is extended to the transmission problems and models of fracture, where localisation and edge waves occur. The theoretical derivations are accompanied with  numerical simulations and illustrations. 
 \end{abstract}
\section{Introduction}

We introduce a novel concept of a multi-scale shield/filter, which couples pressure waves and rotational motion in an elastic lattice. 
Such a structure  incorporates  high-contrast tilted resonators, and their dynamic response is linked to the rotational wave forms.

The interest in elastic waves in chiral media is high, as reflected by the series of papers on micro-structured media, which incorporate active gyroscopes \cite{Brun_2012,Carta_2014, Carta_2017} and \cite{Wang_2015,Susstrunk_2015,Huber_2016}. Waves in such periodic structures possess fascinating, sometimes counter-intuitive, properties. These include filtering, polarisation, as well as directional preference and/or localisation.

 The present paper, in contrast with \cite{Brun_2012,Carta_2014, Carta_2017}, deals with the lattice that does not include any active chiral mechanical elements, like gyroscopic inclusions or a gyroscopic foundation.  However, the geometry of the multi-structure considered here is chiral, and this, in turn, contributes to the coupling between the pressure and shear waves, which is supported by the lattice.  The Bloch-Floquet waves  in doubly periodic structures with tilted lattice resonators, and their dispersion properties, were studied in \cite{Tallarico-2016}. Other geometrically chiral lattices were studied in  \cite{Spadoni_WM_46_435_2009,Bigoni_PRB_87_174303_2013,Spadoni_2012,Liu_2011,Liu_2012} in the continuum approximation. 
 When dealing with effective properties of periodic media, high-frequency homogenisation techniques \cite{Craster_1_2010, Craster_2_2010, Colquitt_2015, Movchan_2013} can be used. 
 
 The notion of the multi-scale multi-structure \cite{kozlov1999fields} was used in \cite{Bigoni_PRB_87_174303_2013} to approximate the frequencies of standing waves of a multi-scale periodic structure with resonators, consisting of discs connected with the ambient medium by thin ligaments. In particular, the issue of degeneracies was noted for configurations of resonators with special inclinations of the thin ligaments. 
 
 The influence  of the micro-structure on a dynamic crack in a lattice was discussed in \cite{colquitt2012trapping,carta2013crack, trevisan_2016}.
 For a transient propagating crack, the crack edge emanates waves, which interact with the ambient medium. Even in subsonic regimes, the problem of a crack advancing in a micro-structured solid is a challenge. Analytical approaches applicable to cracks propagating at an average constant speed  are presented in \cite{slepyan2012models}.\color{black}

We draw the attention of the reader to the papers \cite{Wang_2015,Susstrunk_2015,Huber_2016}, which addressed formation of unidirectional edge waves in active chiral elastic systems by achieving time-reversal symmetry breaking.

In the present work, we give a special attention to micro-structured solids containing cracks, and we show how a coating, built of a tilted resonator lattice,  can absorb vibrations, or otherwise can channel the energy away from the crack tip.\color{black}

An adaptive finite element computation has been performed to model a transient propagation of a crack inside a channel of the micro-structured material. The earlier work \cite{trevisan_2016} has addressed the question of a transient advance of a crack subjected to a dynamic load. The influence of a geometrically chiral multi-scale lattice on the field around the crack is demonstrated in the present paper. 

An additional focus of this paper is on the effect of geometric chirality to the edge waves propagating along structured interfaces. In this context, we would like to mention the earlier work \cite{Joseph_2013} where asymptotics for elastic waves propagating along line defects in triangular and square lattices were investigated. Here we analyse waves around a ``coated'' crack, where the coating is introduced as a multi-scale structure of tilted resonators. 
We show examples of dynamic localisation and edge waves.

The structure of the paper is as follows. The formulation of the problem and an outline of the dispersion properties of the Bloch-Floquet waves in a lattice with tilted resonators are included in Section \ref{Bloch_section}. Wave localisation and edge states are discussed in Section \ref{sec:EW_Loc}. In Section \ref{sec:coated_crack}, we model a crack in a triangular lattice,  surrounded by a structured coating containing tilted resonators. 
In section \ref{sec:crack_trans}, we study an edge crack sandwiched between two strips of resonators and  subjected to a pulsating  thermal load. The advance of  the crack is  studied  in the transient regime. In section \ref{sec:conclusion} we draw our main conclusions.

\begin{figure}[h!]
	\includegraphics[width=0.5\textwidth]{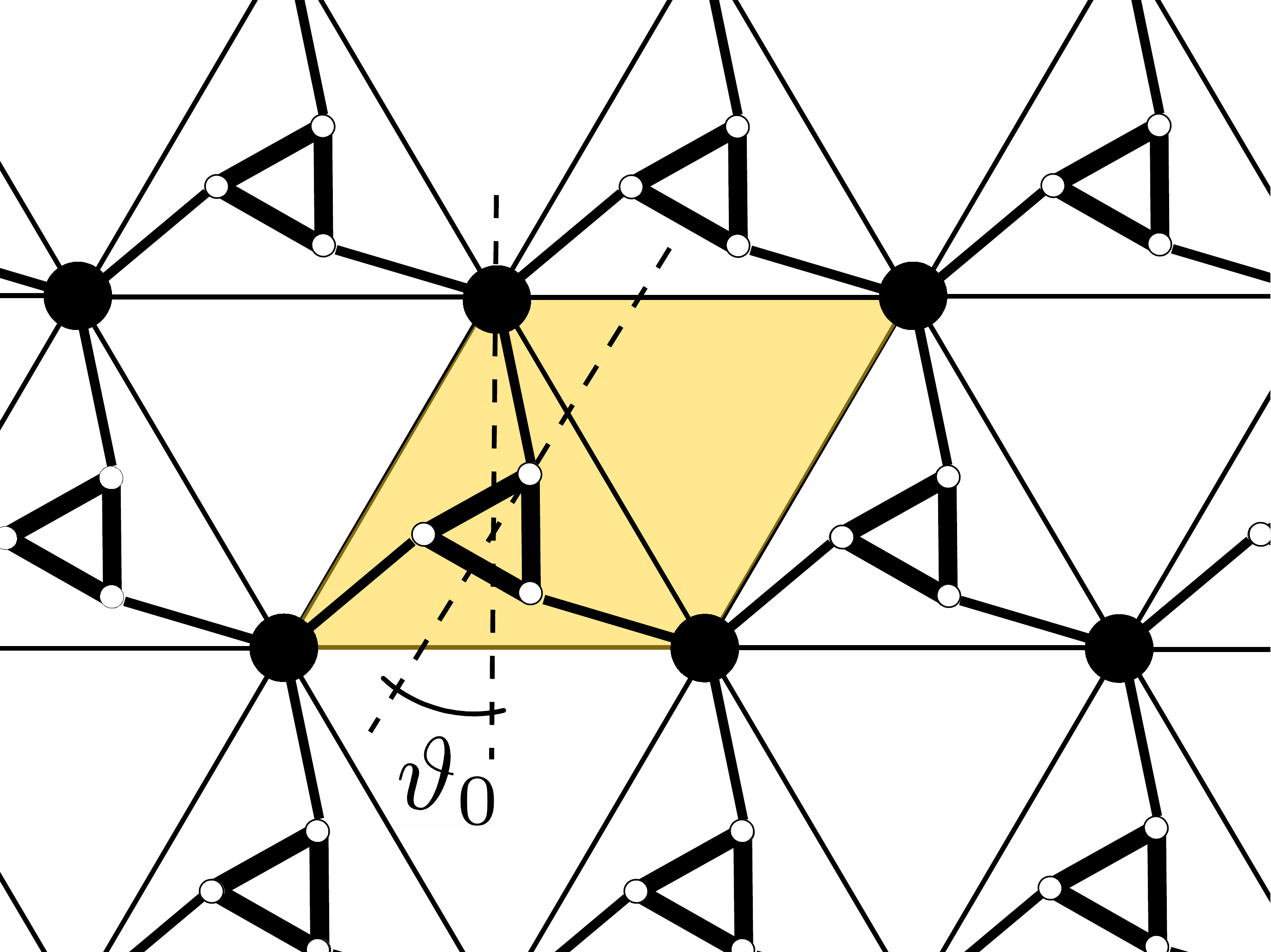}\hfill
	\centering 
	\includegraphics[width=0.5\textwidth]{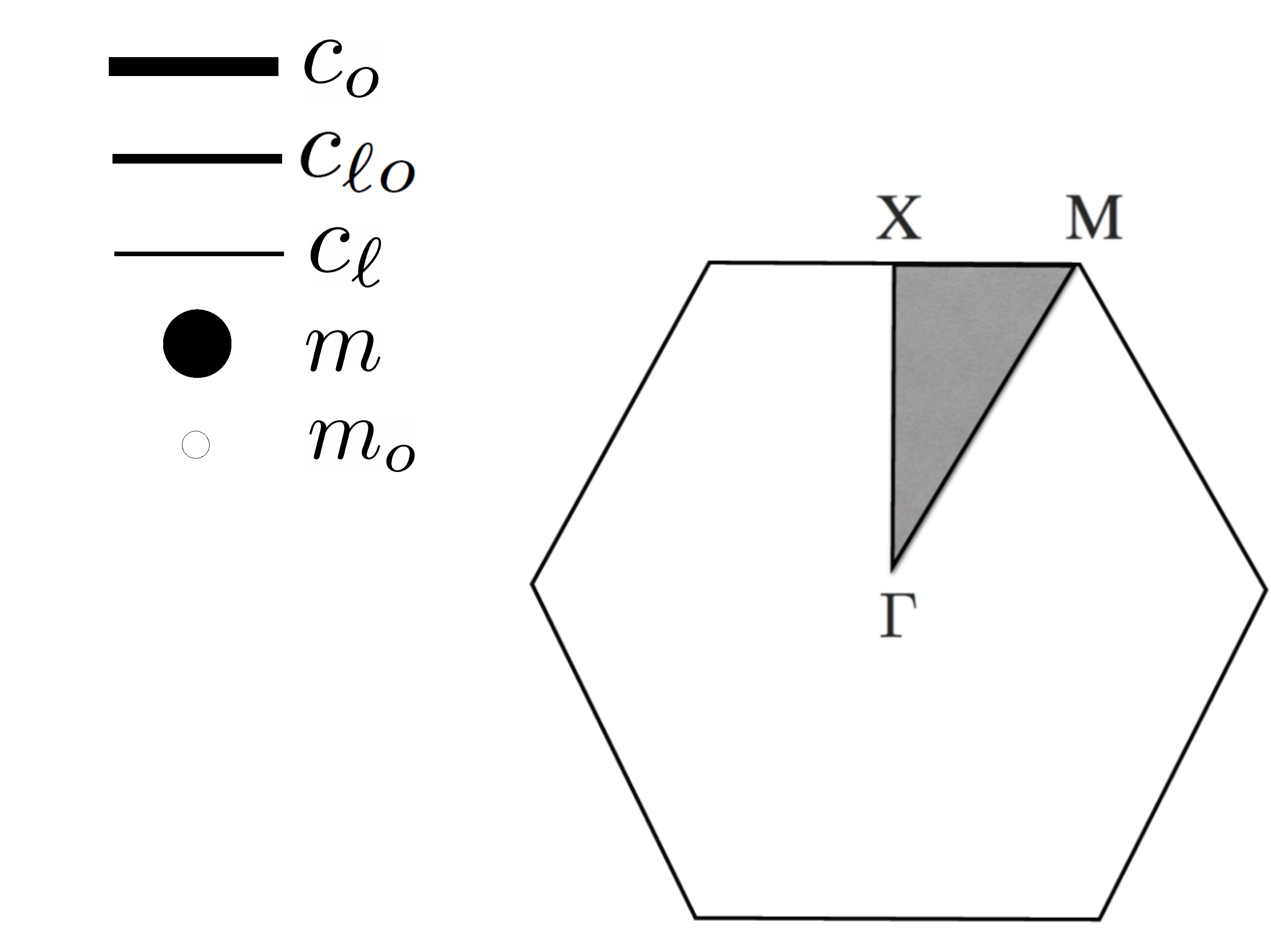}\hfill
	$$\rm \,\,\,\,\,\,\,\,\,\,\,\,\,\,\,\,\,\,\,\,\,\,\,\,\,\,\,\,\,\,\,\,\,\,\,\,\,\,\,\,\,\,\,\,\,\,\,\,\,\,\,\,\,\,\,\,\,\,\,\,\,\,\,\,\,\,\,\,\,\,\,\,\,\,\,\,\,\,\,\,\,\,\,\,\,\,\,\,\,\,\,\,\,(a)\,\,\,\,\,\,\,\,\,\,\,\,\,\,\,\,\,\,\,\,\,\,\,\,\,\,\,\,\,\,\,\,\,\,\,\,\,\,\,\,\,\,\,\,\,\,\,\,\,\,\,\,\,\,\,\,\,\,\,\,\,\,\,\,\,\,\,\,\,\,\,\,\,\,\,\,\,\,\,\,\,\,\,\,\,\,\,\,\,\,\,\,\,\,\,\,\,\,\,\,\,(b)$$
	\centering
	\caption{\label{fig:system_t}Panel (a): A schematic representation of the triangular elastic lattice  containing resonators, tilted by an angle $\vartheta_0$; the unit cell of the lattice is highlighted in yellow. 
		Panel (b): The first Brillouin zone for the triangular lattice and irreducible fraction (grey shaded  region).}
\end{figure}
\section{Bloch-Floquet waves in a triangular lattice with tilted resonators} \label{Bloch_section}
In this section, we 
refer to the earlier paper \cite{Tallarico-2016}, and give an outline describing the propagation of Bloch-Floquet waves in a triangular lattice with tilted rotational resonators. A schematic representation of the triangular lattice with resonators (TLR) is given in Fig. \ref{fig:system_t}(a). 
Here we demonstrate that the Bloch-Floquet frequency dispersion surfaces for the TLR can exhibit Dirac-like dispersion. Dirac-like dispersion arises from the triple-degeneracy of two conical bands and one flat band, as also stated in
\cite{mei2012first}. In contrast, the pure Dirac dispersion is represented by a conical surface, incorporating two cones  above and below the  common vertex, called the ``Dirac point". Such dispersion surfaces are observed, for example,  for lattices of high order of symmetry, such as graphene. Dirac-like dispersion can be achieved via the fine tuning of the unit cell's eigenvalues in a plethora of phononic and photonic metamaterials. \color{black} 
Dirac-like phononic  lattices remain highly attractive because of their interesting physical properties: dynamic neutrality has recently been observed in a  platonic crystal \cite{haslinger2016} and \cite{smith2014double}.
 Perfect transmission and tunnelling were reported  
 in \cite{li2015double} which focussed on photonic crystal governed by the Helmholtz wave equation and exhibiting Dirac-like dispersion.\color{black}
\subsection{Governing equations}
We consider an elastic triangular  lattice (TL) containing tilted rotational resonators, as the one represented in Fig. \ref{fig:system_t}(a). Point-wise masses  $m$ (black full circles) are considered at the triangular lattice nodes in Fig. \ref{fig:system_t}(a), the lattice vectors being
\begin{equation}\label{eq:t_unit_vects}
{\bm t}_1=
\begin{pmatrix}
1\\
0
\end{pmatrix} L\,\,\,\,{\rm and }\,\,\,\,
{\bm t}_2=
\begin{pmatrix}
1\\
\sqrt{3}
\end{pmatrix}\frac{L}{2},
\end{equation}
where $L$ is the distance between nearest neighbours. Fig. \ref{fig:system_t}(b) shows the first Brillouin zone of the TL, together with its irreducible part (grey area). The high-symmetry points are 
\begin{equation}\label{eq:GammaMX}
{\rm \Gamma}= 
\begin{pmatrix}
0 \\
0 \\
\end{pmatrix},
\,\,\,\,\,
{\rm M}= 
\frac{2\pi}{\sqrt{3}L}\begin{pmatrix}
1/\sqrt{3}\\
1\\
\end{pmatrix}
\,\,\,{\rm and}\,\,\,
{\rm X}= 
\frac{2\pi}{\sqrt{3}L}\begin{pmatrix}
0\\
1\\
\end{pmatrix}.
\end{equation}
 The nodal points of the lattice whose mass is  $m$, are linked to each other by non-flexible, massless, extensible rods (thin lines) of longitudinal stiffness $c_\ell$. The unit cell of the lattice (semitransparent yellow region in Fig. \ref{fig:system_t}(a)) contains a resonator, an equilateral triangle of side $\ell$ with point masses $m_o$ attached to its vertices (empty circles in Fig. \ref{fig:system_t}(a)). The vertices of the resonators are linked to the nodal points of the TL by non-flexible, extensible rods of longitudinal stiffness $c_{\ell o}$  (medium thickness black lines in Fig. \ref{fig:system_t}(a)). In this paper the resonators are assumed to be rigid, $\emph{i.e.}$ the longitudinal stiffness $c_o$ of the links connecting the vertices of the resonators is such that $c_o/c_{\ell o}\rightarrow+\infty$ and $c_o/c_{\ell}\rightarrow+\infty$. The resonators are tilted with respect to the external triangular lattice by an angle $\vartheta_0$, marked in Fig. \ref{fig:system_t}(a). 

We give now some geometric definitions useful to represent the dispersion equation for the triangular lattice with resonators. We denote by $\tilde{\bf{b}}_i$, $i=\{1,2,3\}$, the position vector  of the $i^{\rm th }$ mass relative to the centre of mass $\tilde{\bm r}_{\rm cm}=L/2~( 1, 1/\sqrt{3})^{\rm T}$, where ``T'' denotes transposition. The explicit expression is
\begin{equation}\label{eq:b_i_eq}
\tilde{\bf{b}}_{i}=b \hat{\cal R}_i \tilde{\pmb{\beta}}_{1}=
b \hat{\cal R}_i\begin{pmatrix}
\sin \vartheta_0\\
\cos\vartheta_0
\end{pmatrix},\,\,\,\,\,{\rm with}\,\,\,\,\,\hat{\cal R}_i=\left.\hat{\cal R}_{\vartheta}\right|_{\vartheta=2\pi(i-1)/3},~~~ i=\{ 1,2,3\},
\end{equation} 
where  $\vartheta_{0}$ is the tilting angle, $b=\ell/\sqrt{3}$, and
\begin{equation}\label{eq:r_matrix}
\hat{\cal R}_{\vartheta}=
\begin{pmatrix}
\cos{\vartheta} & \sin{\vartheta} \\
-\sin{\vartheta} & \cos{\vartheta} \\
\end{pmatrix},
\end{equation}
is the clockwise rotation matrix. The vector linking the triangular lattice to the $i^{\rm th}$ mass of the resonator in the reference cell ${\bm n}={\bm 0}$ is
\begin{equation}\label{eq:alpha_i}
\tilde{\bm \alpha}_i=\hat{\cal R}_i\tilde{\pmb{\alpha}}_{1}, ~~~ i=\{ 1,2,3\},
\,\,\,\,{\rm with}\,\,\,\,\tilde{\pmb{\alpha}}_{1}=\bm{t}_2 - \tilde{\bm r}_{\rm cm}-\tilde{\bm b}_1= 
\begin{pmatrix}
b \sin \vartheta_0 \\
-(B-b\cos\vartheta_0 )
\end{pmatrix},
\end{equation}
where $B=L/\sqrt{3}$, $b$ has been introduced in Eq. \eqref{eq:b_i_eq} and the matrix $\hat{\cal R}_i$ is given in Eq. (\ref{eq:b_i_eq}). Given the set of vectors  (\ref{eq:t_unit_vects}) and (\ref{eq:alpha_i}), we introduce the corresponding projector matrices 
\begin{align}\label{eq:projectors}
\hat{\tau}_1&=\frac{1}{L^2} {\bm t}_1 {\bm t}_1^{\rm T},\,\,\,\,\,\,
\hat{\tau}_2=\frac{1}{L^2} {\bm t}_2 {\bm t}_2^{\rm T},\,\,\,\,\,\,
\hat{\tau}_3=\frac{1}{L^2}\left( {\bm t}_1- {\bm t}_2\right) \left( {\bm t}_1- {\bm t}_2\right)^{\rm T}, \,\,\,\,\,\,\nonumber\\
\hat{\Pi}_i &=\frac{1}{\ell_r^2} \tilde{\bm{\alpha}}_i\tilde{\bm{\alpha}}_i^{\rm T},~~~i=\{1,2,3\},~~~{\rm with}~~\ell_r = ||\tilde{\bm \alpha}_i||= \frac{1}{\sqrt{3}} \sqrt{L^2+\ell^2-2\ell L\cos(\vartheta_0)}.
\end{align}
The notation ${\bm v}{\bm u}^{\rm T}$ in Eqs (\ref{eq:projectors}) is used to denote the dyadic product ${\bm v}\otimes{\bm u}$ of two vectors ${\bm u}$ and ${\bm v}$.

We consider time-harmonic elastic Bloch-Floquet waves propagating  through the lattice.  Following \cite{Tallarico-2016},  the Bloch-Floquet displacement wave's amplitude with Bloch-Floquet wave vector ${\bm k}$ is 
\begin{equation}\label{eq:bloch_ampltitude}
{\bm U}_{\bm k}=\begin{pmatrix}
{\bm u}_0^{\rm T}({\bm k}),&{\bm u}^{\rm T}_{\rm cm}({\bm k}),&\vartheta({\bm k})
\end{pmatrix}^{\rm T},
\end{equation}
where the vectors quantities ${\bm u}_0^{\rm T}({\bm k})$ and ${\bm u}^{\rm T}_{\rm cm}({\bm k})$ are the in-plane displacements of the TL nodal points and of the centre of mass of the resonators, respectively.  In Eq. \eqref{eq:bloch_ampltitude}, $\vartheta({\bm k})$ represents the angular displacement with respect to the equilibrium $\vartheta_0$.  In the time-harmonic regime, the equations of motion in the lattice characterised by the displacement \eqref{eq:bloch_ampltitude}, have the matrix form 
\begin{equation}\label{eq:disp_eq}
\left(\hat{\Sigma}_{\bm k}-\omega^2\hat{\cal M}\right){\bm U}_{\bm k}={\bm 0},
\end{equation}
where $\omega$ is the Bloch-Floquet radian frequency and the vector ${\bm U}_{\bm k}$ is given in Eq. \eqref{eq:bloch_ampltitude}.
The inertia matrix which appears in Eq. \eqref{eq:disp_eq} is 
\begin{equation}\label{eq:mass_BF_rigid}
\hat{\cal M}={\rm diag}(m,m,M,M,I),
\end{equation}
where $M=3m_o$ is the total mass of the resonator, $I=m_o\ell^2$ is its moment of inertia, and $m$ is the mass of the nodal points of the triangular lattice. In \cite{Tallarico-2016} it has been shown that  the stiffness matrix in Eq. \eqref{eq:disp_eq} is
\begin{align}\label{eq:Sigma_k}
\hat{\Sigma}_{\bm k}&=\begin{pmatrix}
\hat{\Sigma}_{0,0}({\bm k})& \hat{\Sigma}_{0,{\rm cm}}({\bm k})&{\bm \Sigma}_{0,\vartheta}({\bm k})\\
\\
\hat{\Sigma}^{\dagger}_{0,{\rm cm}}({\bm k}) & \hat{\Sigma}_{{\rm cm},{\rm cm}} &  {\bm \Sigma}_{{\rm cm},\vartheta}\\
\\
{\bm \Sigma}^{\dagger}_{0,\vartheta}({\bm k}) & {\bm \Sigma}^{\dagger}_{{\rm cm},\vartheta}& {\Sigma}_{\vartheta,\vartheta}
\end{pmatrix},\\
&=\sum_{i=1}^3\begin{pmatrix}
-2 c_\ell(\cos({\bm k}\cdot{\bm t}_i)-1)\hat{\tau}_i+c_{\ell o}\hat{\Pi}_i& -c_{\ell o}\varphi_{i}({\bm k})\hat{\Pi}_i&-c_{\ell o} \varphi_{i}({\bm k})\hat{\Pi}_i\hat{\cal R}'_{i}\tilde{\bm b}_1\\
\\
-c_{\ell o} \varphi^*_{i}({\bm k} )\hat{\Pi}_i & c_{\ell o}\hat{\Pi}_i &  c_{\ell o} \hat{\Pi}_i\hat{\cal R}'_{i}\tilde{\bm b}_1\\
\\
-c_{\ell o} (\varphi_{i}({\bm k})\hat{\Pi}_i\hat{\cal R}'_{i}\tilde{\bm b}_1)^\dagger&c_{\ell o}(\hat{\Pi}_i\hat{\cal R}'_{i}\tilde{\bm b}_1)^{\dagger}& -c_{\ell o}\tilde{\bm b}_1^{\rm T}\cdot (\hat{\cal R}_i'\hat{\Pi}_i\hat{\cal R}_i'\tilde{\bm b}_1)
\end{pmatrix},
\end{align}
where  $\varphi_1({\bm k})={\rm exp}(-{\bm k}\cdot{\bm t}_2)$, $\varphi_2({\bm k})={\rm exp}(-{\bm k}\cdot{\bm t}_1)$, $\varphi_3({\bm k})=1$ and $\hat{\cal R}'_i=d/{d\vartheta}\left.\left(\hat{\cal R}_{\vartheta}\right)\right|_{\vartheta=2\pi(i-1)/3}$.
Consider the  $3\times3$ block independent of ${\bm k}$ which appears in Eq. \eqref{eq:Sigma_k}. We observe that
\begin{equation}\label{eq:sigma_single}
\sigma=
\begin{pmatrix}
\hat{\Sigma}_{{\rm cm},{\rm cm}} &  {\bm \Sigma}_{{\rm cm},\vartheta} \\
\\
{\bm \Sigma}^{\dagger}_{{\rm cm},\vartheta}& {\Sigma}_{\vartheta,\vartheta}
\end{pmatrix}
=\begin{pmatrix}
3c_{\ell o} /2~\hat{I}_{2 \times 2} &  {\bm 0} \\
\\
{\bm 0}^{\rm T}& c_{\ell o} \ell^2\sin^2\vartheta_0/(1+\ell^2/L^2-2 \ell/L \cos(\vartheta_0))
\\
\end{pmatrix}, 
\end{equation}
where $\hat{I}_{2\times2}$ is the $2\times2$ identity matrix. The diagonal  matrix \eqref{eq:sigma_single} is the stiffness matrix for a single resonator for which the natural frequencies squared are \cite{Tallarico-2016}
\begin{equation}\label{eq:freq_single}
\Omega_{\rm cm}=\frac{3}{2} \frac{c_{\ell o}}{M},\,\, {\rm and}\,\,\,\Omega_{\vartheta}=\frac{c_{\ell o}\ell^2}{I}\frac{ \sin^2\vartheta_0}{1+\ell^2/L^2-2\ell/L\cos\vartheta_0}. 
\end{equation}
In Eq. \eqref{eq:freq_single},  $\Omega_{\rm cm}$ is the frequency of oscillation of the centre of mass of a single resonator, whereas the frequency $\Omega_{\vartheta}$ describes the harmonic rotation of the resonator.  
\begin{figure}[t!]
	\centering
	\includegraphics[width=0.5\textwidth]{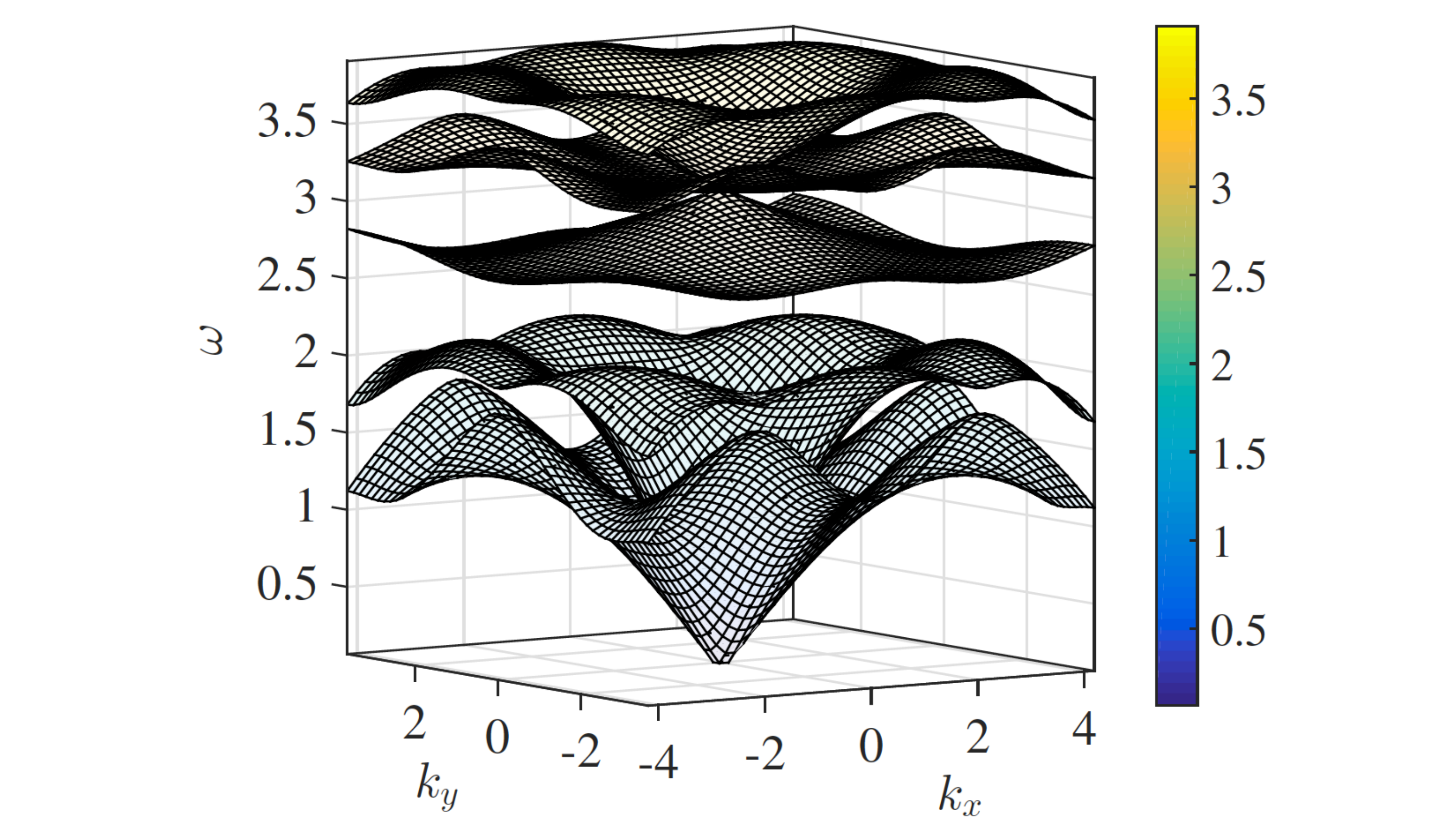}\\
	\caption{\label{fig:disp_DC_3D} The Bloch-Floquet dispersion surfaces for a triangular lattice with resonators whose lattice parameters are  listed in set 1 of Table \ref{tab:parameters}. The colour scale represents Bloch-Floquet frequencies $\omega$.}
\end{figure}
\begin{figure}[t!]
\centering
\includegraphics[width=0.5\textwidth]{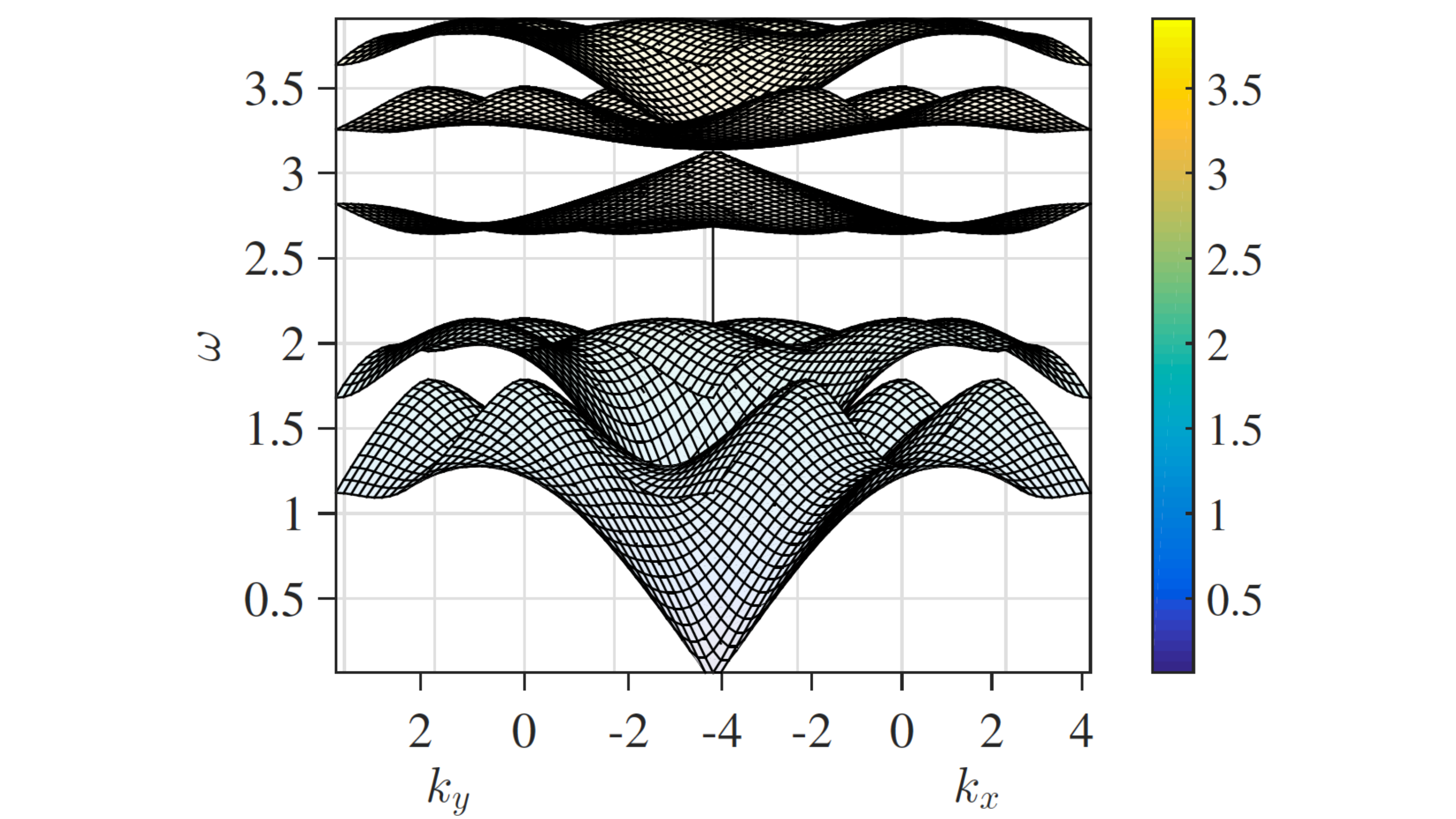}\hfill
\includegraphics[width=0.5\textwidth]{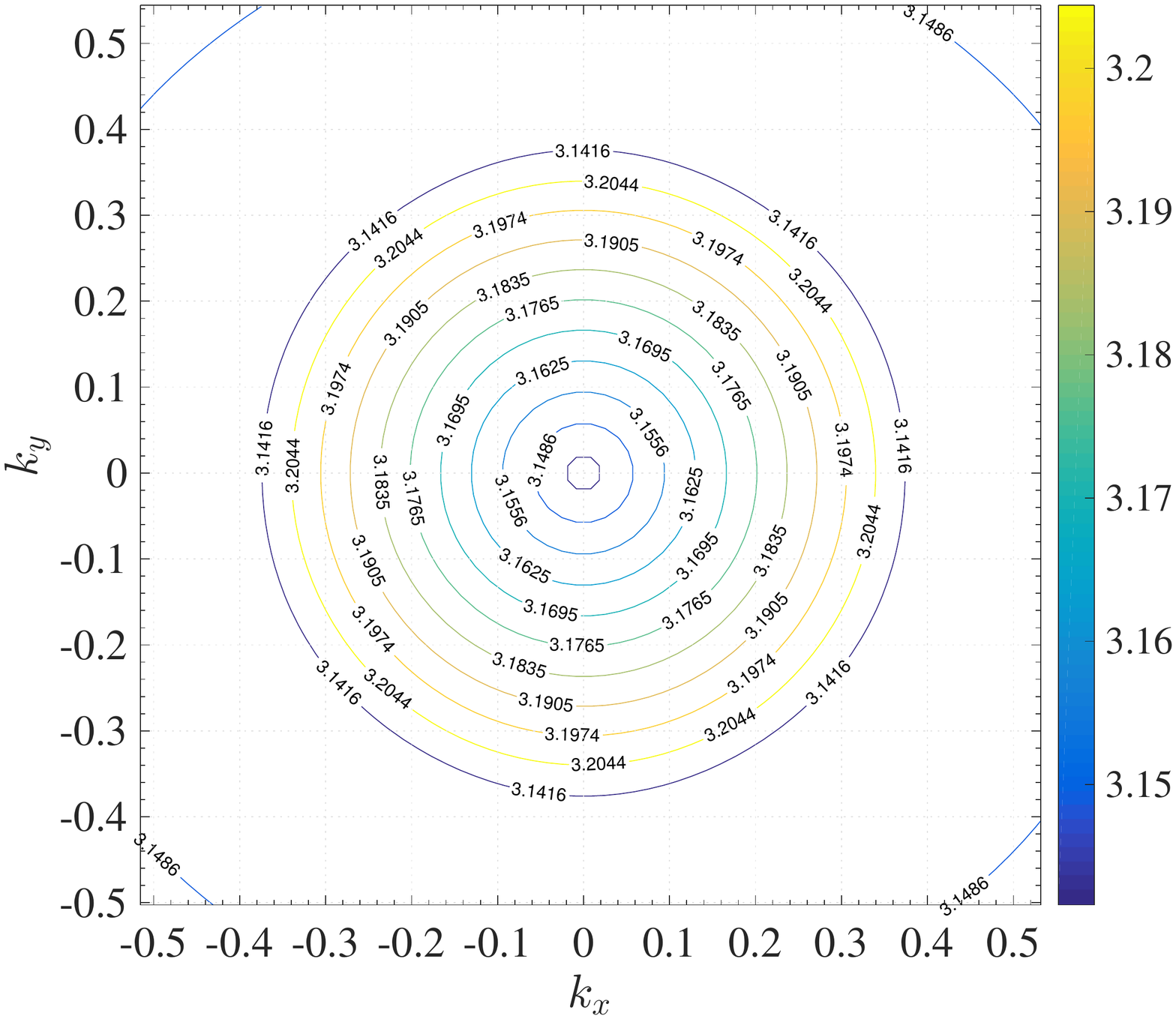}\\
$(a)\,\,\,\,\,\,\,\,\,\,\,\,\,\,\,\,\,\,\,\,\,\,\,\,\,\,\,\,\,\,\,\,\,\,\,\,\,\,\,\,\,\,\,\,\,\,\,\,\,\,\,\,\,\,\,\,\,\,\,\,\,\,\,\,\,\,\,\,\,\,\,\,\,\,\,\,\,\,\,\,\,\,\,\,\,\,\,\,\,\,\,\,\,\,\,\,\,\,\,\,\,\,\,\,\,\,\,\,\,\,\,\,\,\,\,\,\,\,\,\,\,\,\,\,\,\,\,\,\,\,\,(b)$\\
\includegraphics[width=0.5\textwidth]{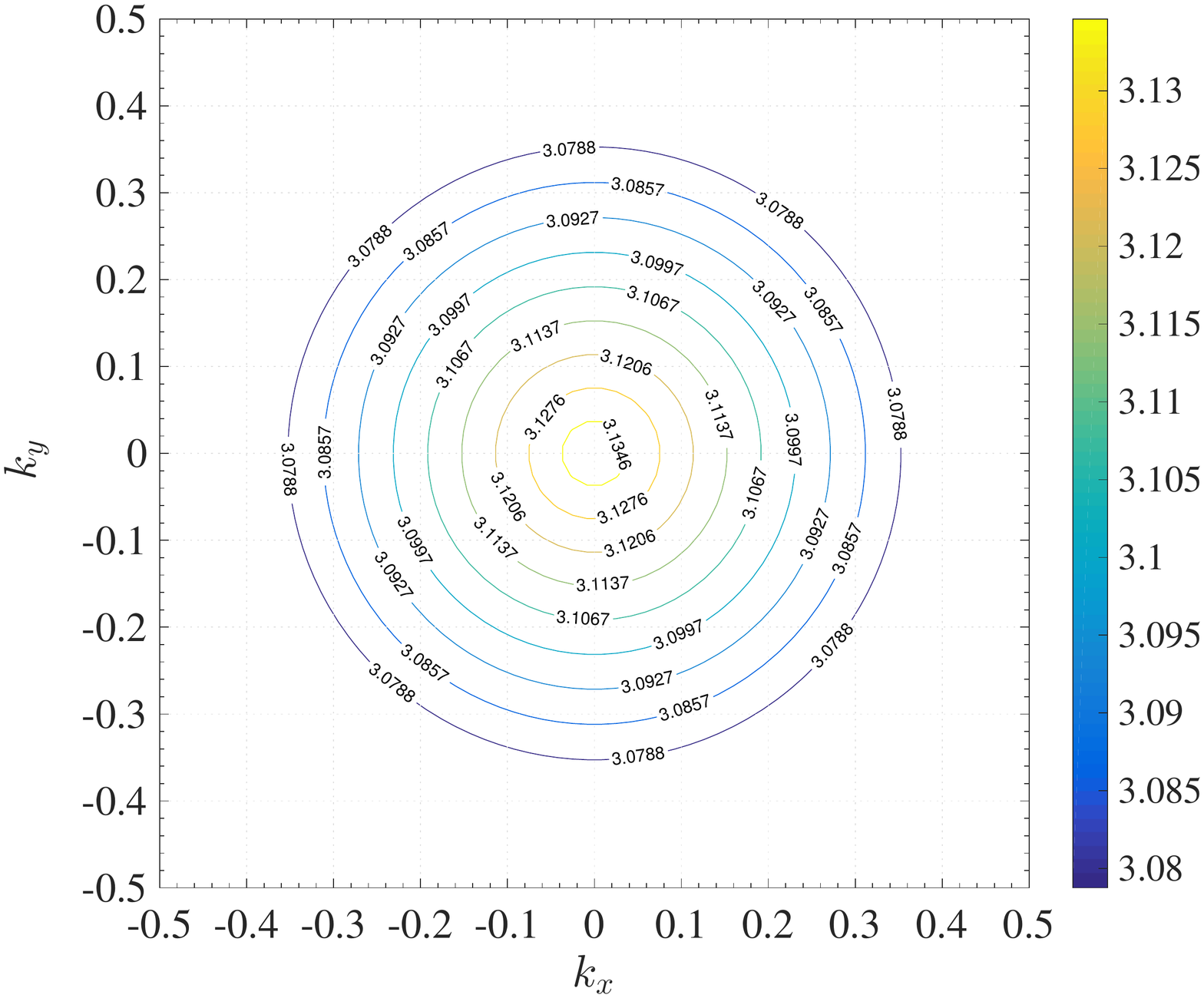}\hfill
\includegraphics[width=0.5\textwidth]{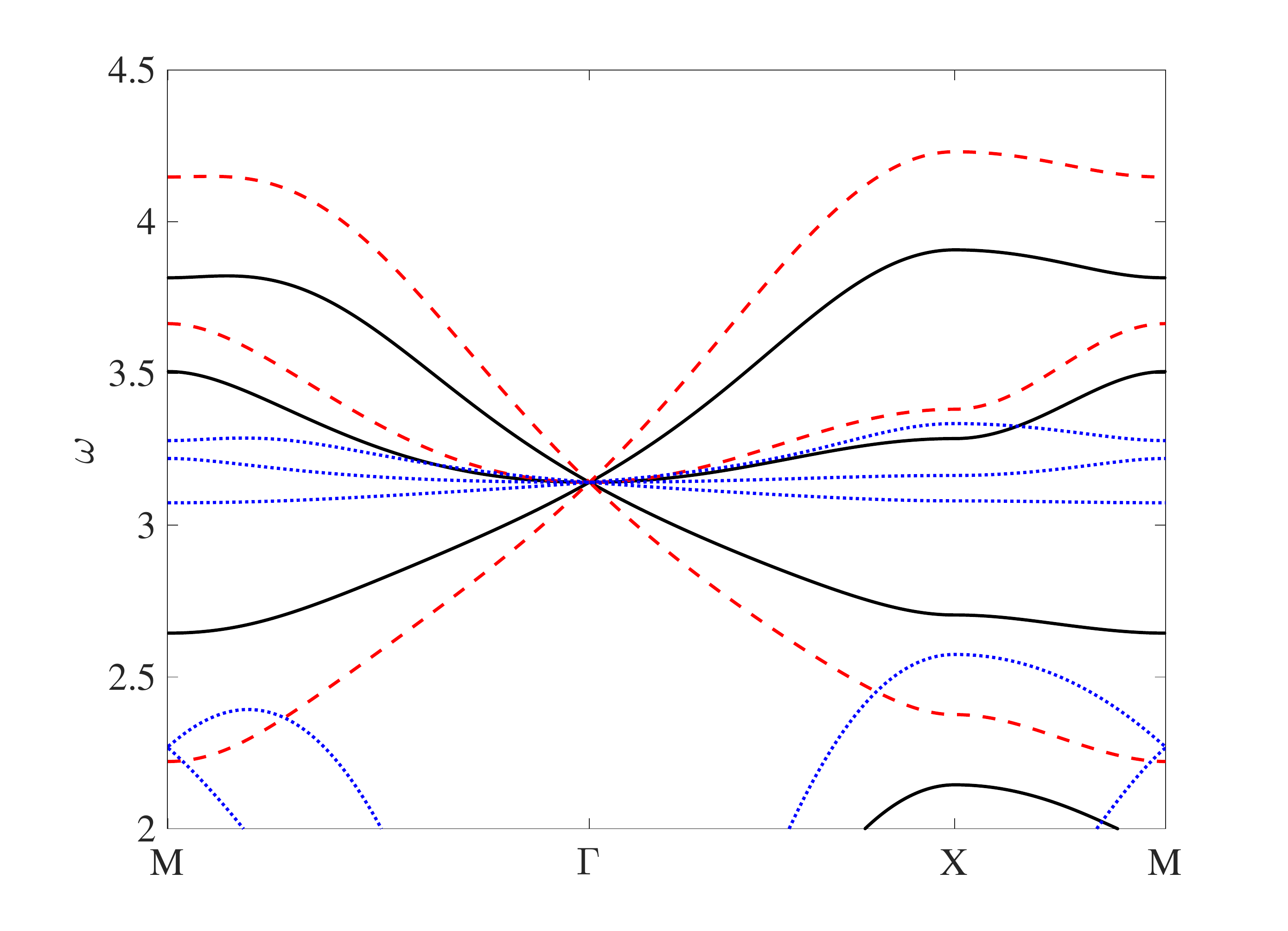}\\
$(c)\,\,\,\,\,\,\,\,\,\,\,\,\,\,\,\,\,\,\,\,\,\,\,\,\,\,\,\,\,\,\,\,\,\,\,\,\,\,\,\,\,\,\,\,\,\,\,\,\,\,\,\,\,\,\,\,\,\,\,\,\,\,\,\,\,\,\,\,\,\,\,\,\,\,\,\,\,\,\,\,\,\,\,\,\,\,\,\,\,\,\,\,\,\,\,\,\,\,\,\,\,\,\,\,\,\,\,\,\,\,\,\,\,\,\,\,\,\,\,\,\,\,\,\,\,\,\,\,\,\,\,(d)$\\
	\caption{\label{fig:disp_DC} In panel (a) a side view of Fig. \ref{fig:disp_DC_3D} is provided. Panels (b) and (c) are  slowness contours of panel (a). The frequencies represented here lie just above (panel (b)) and just below (panel (c))  $\omega=\pi$, corresponding to the Dirac-like point.The colour scale represents Bloch-Floquet frequencies $\omega$. Panel (d) shows the dispersion curves of the  optical modes for three set of lattice parameters. The  solid black lines correspond to the lattice parameters used in panel (a); red dashed lines and blue dotted lines correspond to ``set 2" and ``set 3" in Table \ref{tab:parameters}, respectively. }
\end{figure}
\begin{table}[h]
	\centering
	\begin{tabular}{@{}lllllllr@{}} \toprule
                              &$c_\ell$ & $m$               & $L$ & $\ell$ & $c_{\ell o}$ &$ m_o$       & $\vartheta_{0}$\\ \midrule
		${\rm set~ 1}$&1            &                  0.8& 1     &   0.21  &      1.534     &         0.11   & 0.82\\
		 ${\rm set~ 2}$&1           &                  0.8& 1     &   0.25  &      2.6319       &         0.27   & 1.32\\
		 ${\rm set~ 3}$&1           &                  0.8& 1     &     0.1  &      0.2722   &     0.0145   & 0.74\\\bottomrule
	\end{tabular}\caption{\label{tab:parameters}Sets of parameters for selected triangular lattices with resonators whose frequency dispersion (see Fig. \ref{fig:disp_DC}) is Dirac-like at $\omega=\pi$. SI units of measurement are understood.}
\end{table}
\subsection{Triple eigenvalue and Dirac-like dispersion surfaces near ${\bf k}={\bf 0}$ \label{sec:DC}}

The elastic Bloch-Floquet waves in the doubly-periodic structure of tilted resonators have interesting dispersion properties shown in  Fig. \ref{fig:disp_DC_3D}. A special feature is the Dirac-like cone with the vertex corresponding to ${\bf k}={\bf 0}$, which is the main focus of this paragraph. Seeking non-trivial solutions for Eq. \eqref{eq:disp_eq} requires
\begin{equation}\label{eq:disp_eq_JMPS}
{\cal D}({\bm k},\omega)={\rm det}\left(\hat{\Sigma}_{\bm k}-\omega^2\hat{\cal M}\right)=0,
\end{equation}  
whose roots $\omega$ \emph{vs} ${\bm k}$ determine the dispersion of  Bloch waves (see \emph{e.g.} Fig. \eqref{fig:disp_DC_3D}).  At ${\bm k}={\bm 0}$, the roots of the fifth-degree in $\Omega=\omega^2$  polynomial equation \eqref{eq:disp_eq_JMPS} can be found in their closed forms.  Introducing the notation $\Omega^{(i)}_{\Gamma}=\left.\Omega^{(i)}_{\bm k}\right|_{{\bm k}={\bm 0}}$, with $i$ the index of the root, we find 
\begin{equation}\label{eq:eigf_gamma}
\Omega^{(1)}_{\Gamma}=0,\,\,\,\Omega^{(2)}_{\Gamma}=\Omega_{\rm cm}\left(1+\frac{3m_o}{m}\right)\,\,\,{\rm and}\,\,\,\Omega^{(3)}_{\Gamma}=\Omega_{\vartheta}.
\end{equation}
where $\Omega_{\rm cm}$ and $\Omega_{\vartheta}$ have been introduced in Eq. \eqref{eq:freq_single}. The first and second eigenvalues in Eqs. \eqref{eq:eigf_gamma} have multiplicity two, and the third one has multiplicity one. 
The geometric conditions 
 \begin{equation}\label{eq:geom_cond}
 0<\frac{\ell}{L}<\frac{1}{2}\,\,\,\,\,\,\,\,\,{\rm and }\,\,\,\,\,\,\,\,\,|\vartheta_0|<\vartheta_{\rm max}\equiv{\rm arcos}\left(\frac{\ell}{L}\right),
 \end{equation}
guarantee that the trusses do not cross each another. We observe that it is possible to obtain a triple eigenvalue corresponding to $\Omega^{(2)}_{\Gamma}=\Omega^{(3)}_{\Gamma}$, if there exists 
\begin{equation}\label{eq:cond_3e}
\bar{m}= \frac{\cos 2\vartheta_0 + \bar{\ell}^2- 2 \bar{\ell}\cos \vartheta_0}{2\bar{\ell}\cos\vartheta_0- \bar{\ell}^2-1}>0,
\end{equation}
with $\bar{m}=3m_o/m$ and $\bar{\ell}=\ell/L$. We observe that 
\begin{equation}\label{eq:mass_positive}
\bar{m}>0  \Longleftrightarrow\,\,\, \cos\vartheta_0-|\sin\vartheta_0|<\bar{\ell}<\cos\vartheta_0+|\sin\vartheta_0|.
\end{equation} The substitution of the  expression \eqref{eq:cond_3e} for $m_o$ into the  Bloch frequencies at $\Gamma$ in Eq. \eqref{eq:eigf_gamma} gives the frequency squared  for the triple eigenvalue 
\begin{equation}\label{eq:eigf_gamma_tr}
\Omega_{\rm \Gamma}^{\rm (te)}=-\frac{3 c_{\ell o} }{m}\frac{\sin^2\vartheta_0}{\bar{\ell}^2-2 \bar{\ell} \cos \vartheta_0+\cos 2 \vartheta_0},
\end{equation}
which is a positive quantity if the condition on $\bar{\ell}$ and $\vartheta_0$ of Eq. \eqref{eq:mass_positive} is satisfied.

Fig. \ref{fig:disp_DC}(a)  represents the frequency dispersion surfaces for a TLR as a function of  a set of Bloch wave vectors  which comprise the first Brillouin zone (see Fig. \ref{fig:system_t}(b)).  The lattice parameters have been chosen in such a way that Eq. \eqref{eq:cond_3e} is satisfied. This implies the occurrence of a triple eigenvalue at  $\Gamma$, as it can be seen by direct inspection of the optical part of the dispersion diagram. Specifically,  we choose $\bar{\ell}=0.21$ and $\vartheta_0=0.82$, which gives $\bar{m}=0.41$. Moreover, we fix $L=c_{\ell}=1$ and $m=0.8$ which influences the maximum frequency of the acoustic modes.  Finally, the choice $c_{\ell o}=1.53$ guarantees that the triple-eigenvalue's frequency is 
\begin{equation}\label{eq:DC_pinning}
\sqrt{\Omega_{\Gamma}^{\rm (te)}}=\pi.
\end{equation}
Figs \ref{fig:disp_DC}(b) and \ref{fig:disp_DC}(c) show the slowness contours of Fig. \ref{fig:disp_DC}(a) around the triple-eigenvalue's frequency $\omega=\pi$. Fig. \ref{fig:disp_DC}(b) (Fig. \ref{fig:disp_DC}(c)) refers to frequencies slightly above (slightly below) $\omega=\pi$.  Figs \ref{fig:disp_DC}(b) and \ref{fig:disp_DC}(c) show that the dispersion in the vicinity of the triple-eigenvalue is isotropic.  In Fig. \ref{fig:disp_DC}(d) we compare along the path $\rm M\Gamma XM$ the optical branches of three different TLRs whose lattice parameters are listed in Table \ref{tab:parameters} . The black solid line refers to the Set 1 in Table \ref{tab:parameters} which has been already used in Fig. \ref{fig:disp_DC}(a). Hence, the dispersion around the triple-eigenvalue's frequency $\omega=\pi$ is linear, suggesting that the triple eigenvalue is a  Dirac-like point.  Other choices of the parameters are possible resulting in different effective group velocities at $\Gamma$. In Fig. \ref{fig:disp_DC}(d) we use the set 2 (red dashed line) and set 3 (blue dotted line) listed in Table \ref{tab:parameters}. The chosen sets of parameters satisfy \eqref{eq:DC_pinning}, which corresponds to the occurrence of a triple eigenvalue at ${\rm \Gamma}$ and $\omega=\pi$. We observe that Dirac-like dispersion is robust over the chosen sets of the lattice parameters. 
\begin{figure}[h!]
	\centering
	\includegraphics[width=0.33\textwidth]{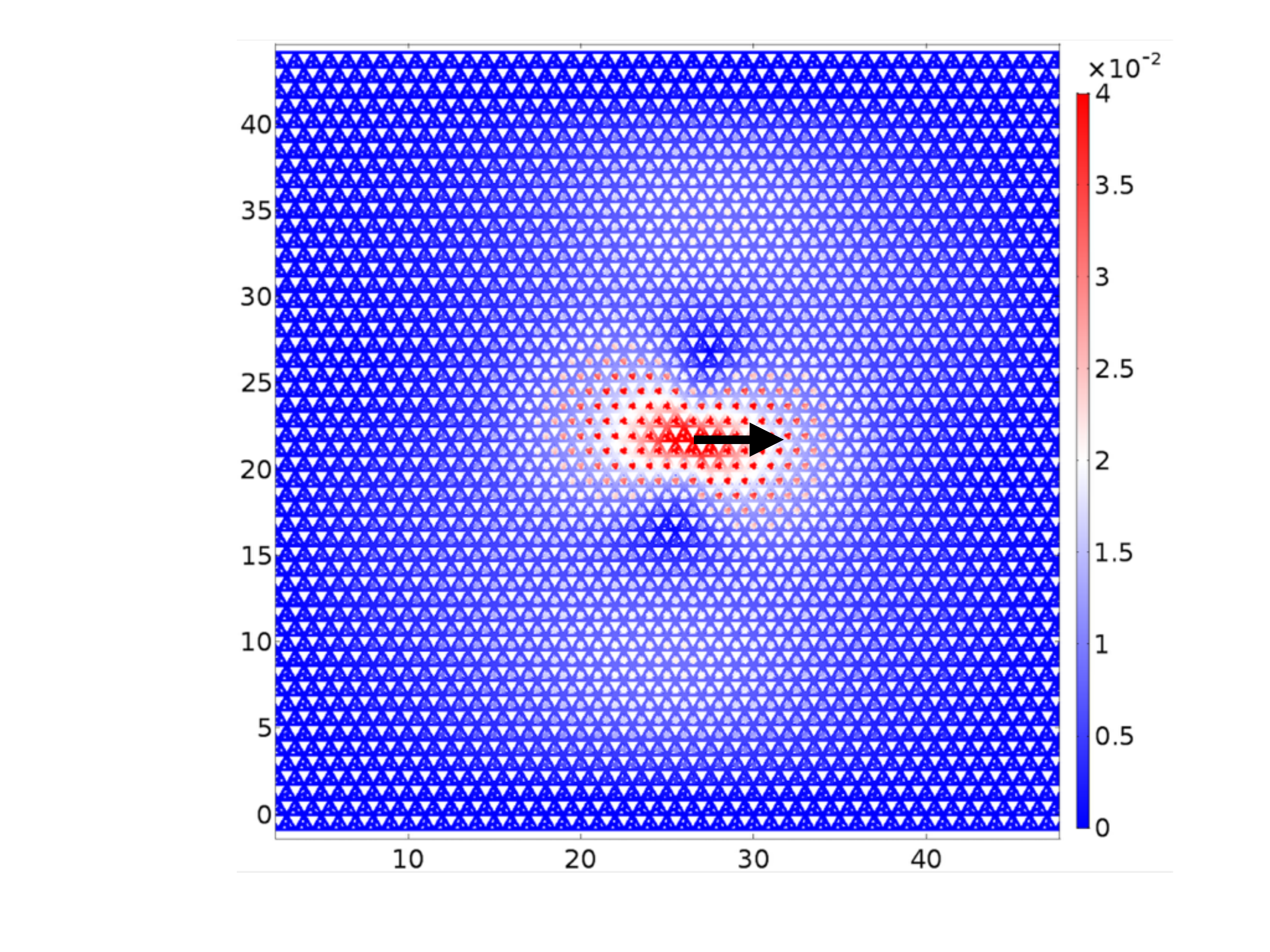}\hfill
	\includegraphics[width=0.33\textwidth]{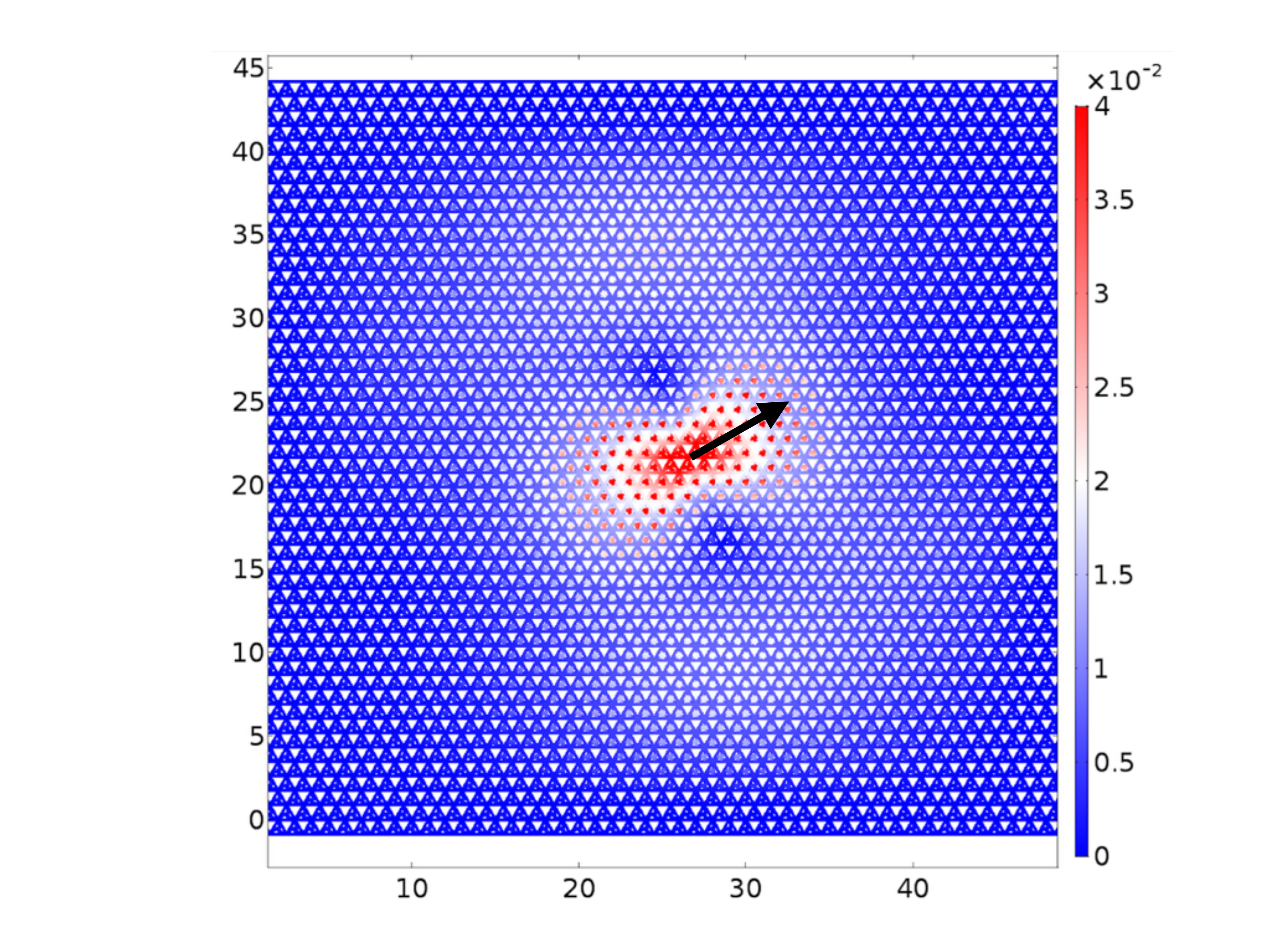}\hfill
	\includegraphics[width=0.33\textwidth]{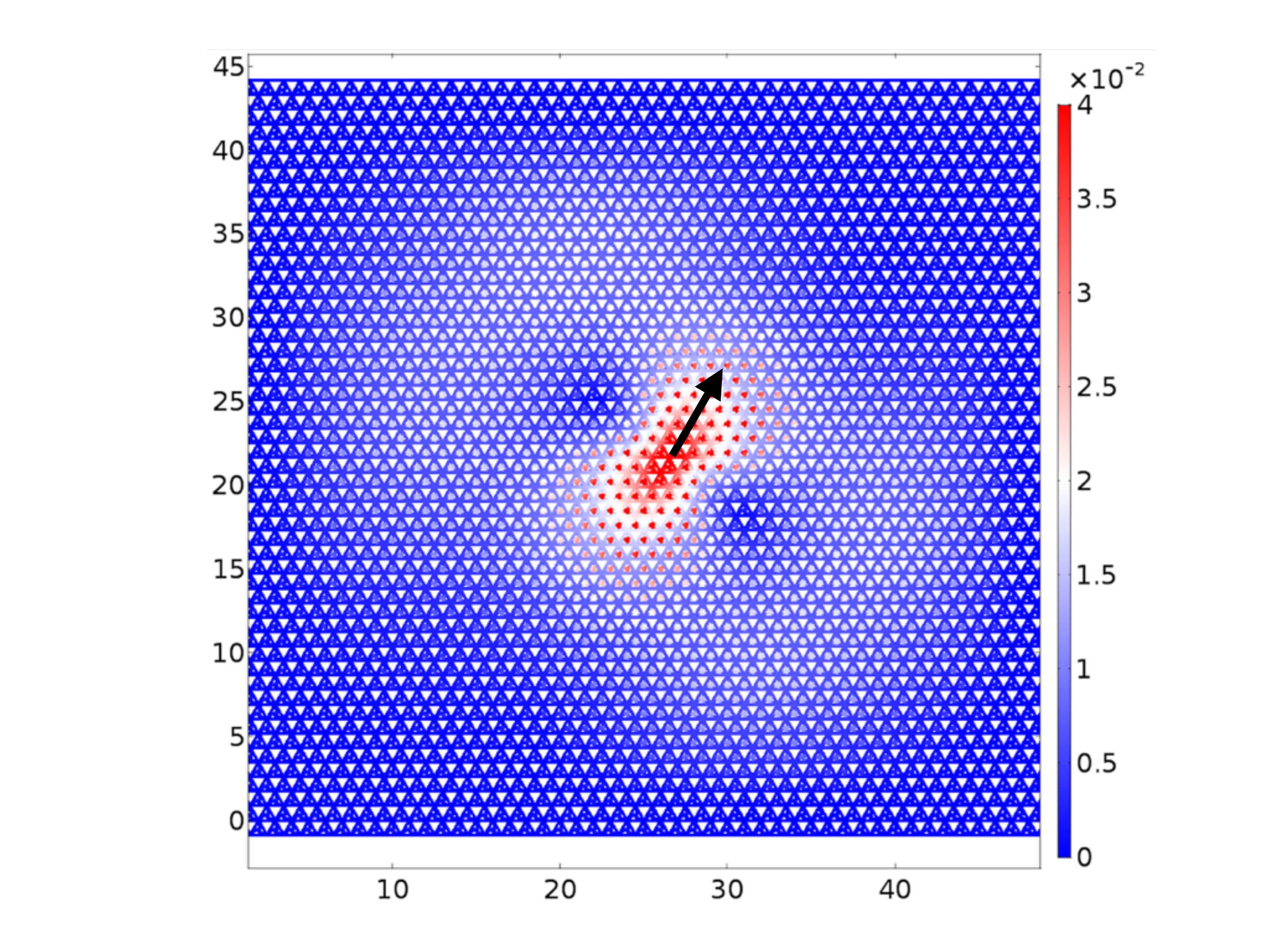}
	$$(a)\,\,\,\,\,\,\,\,\,\,\,\,\,\,\,\,\,\,\,\,\,\,\,\,\,\,\,\,\,\,\,\,\,\,\,\,\,\,\,\,\,\,\,\,\,\,\,\,\,\,\,\,\,\,\,\,\,\,\,\,\,\,\,\,\,\,\,\,\,\,\,\,\,\,\,\,\,\,\,\,(b)\,\,\,\,\,\,\,\,\,\,\,\,\,\,\,\,\,\,\,\,\,\,\,\,\,\,\,\,\,\,\,\,\,\,\,\,\,\,\,\,\,\,\,\,\,\,\,\,\,\,\,\,\,\,\,\,\,\,\,\,\,\,\,\,\,\,\,\,\,\,\,\,\,\,\,\,\,(c)$$
	\includegraphics[width=0.33\textwidth]{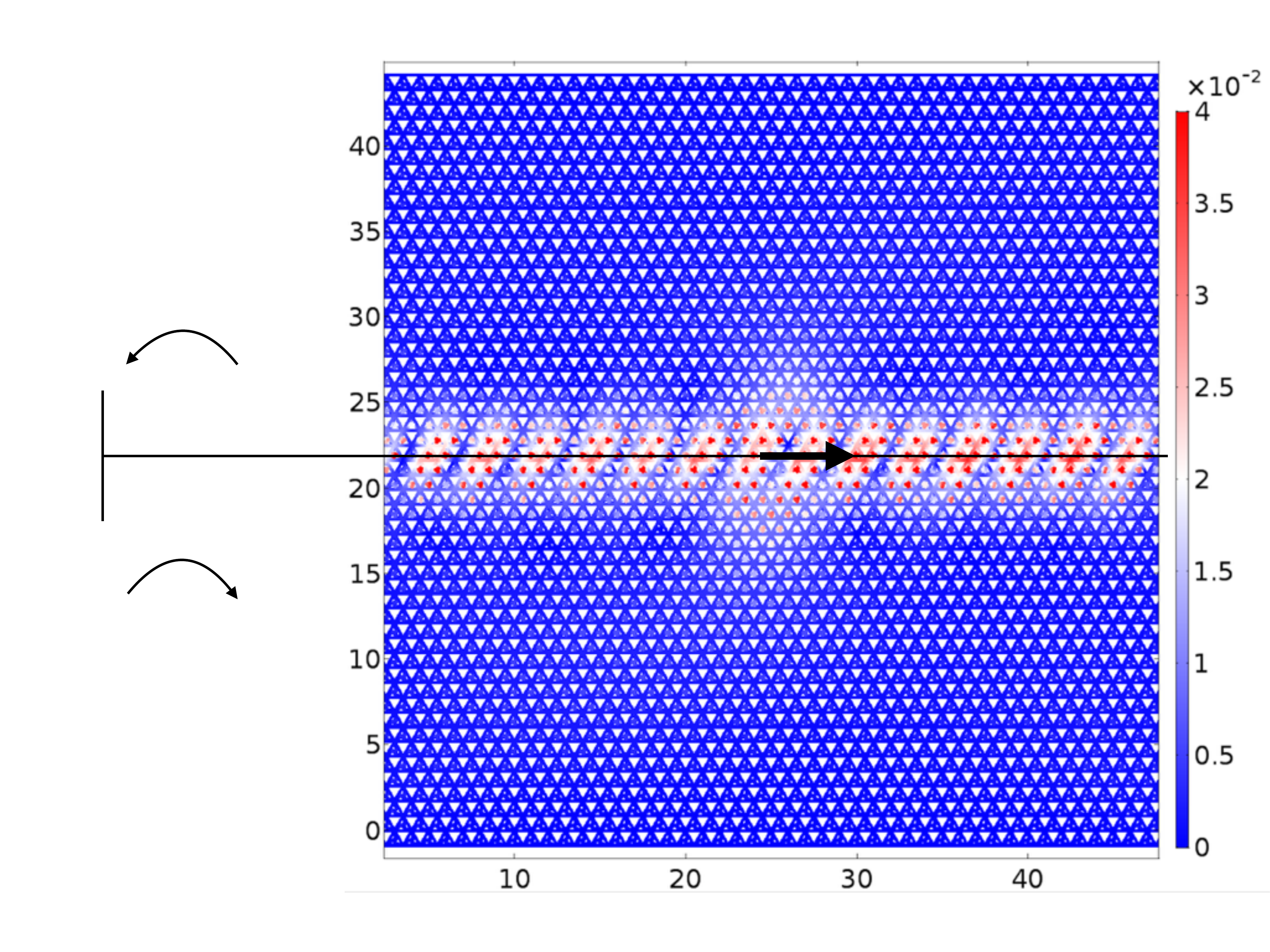}\hfill
	\includegraphics[width=0.33\textwidth]{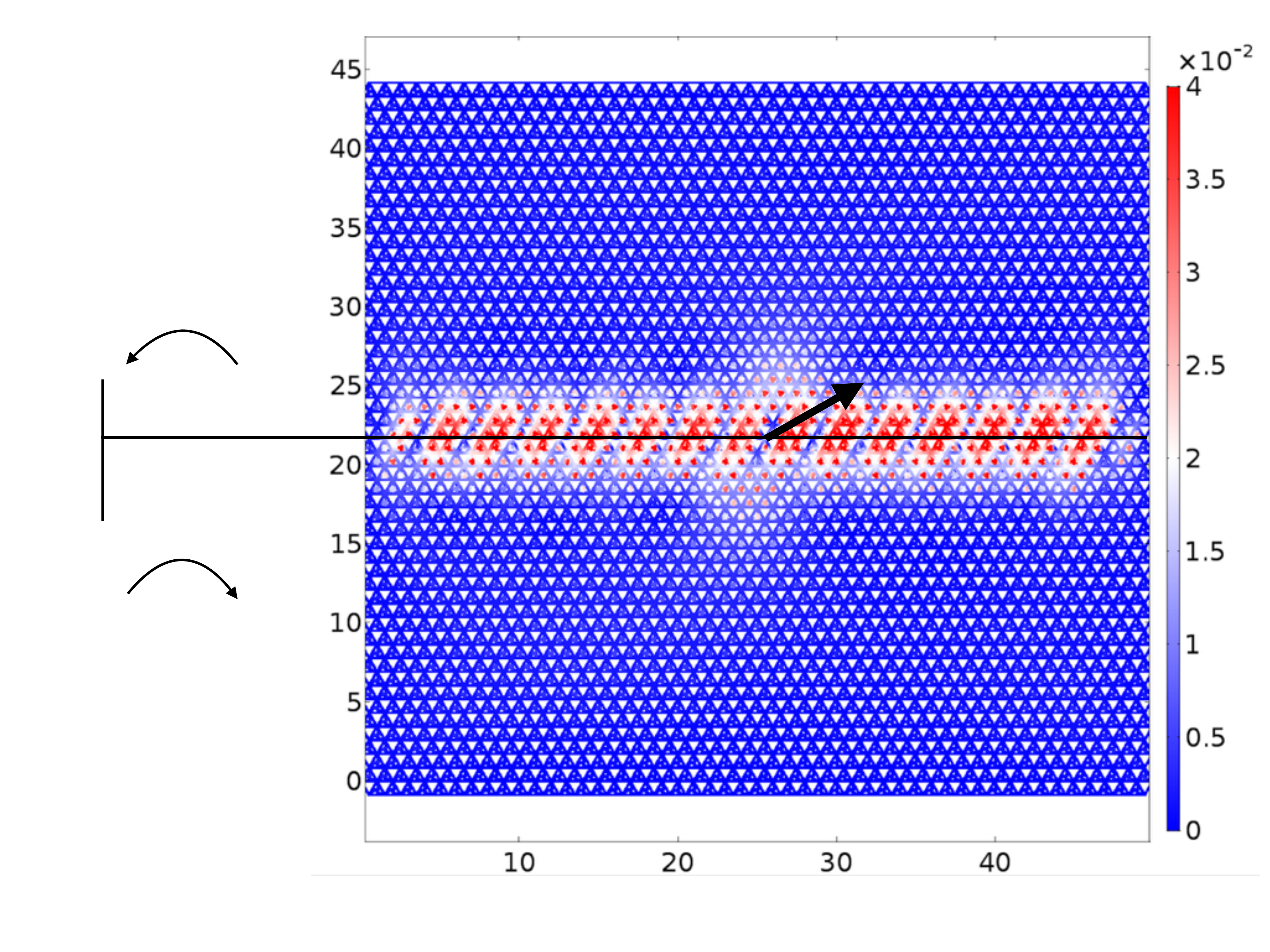}\hfill
	\includegraphics[width=0.33\textwidth]{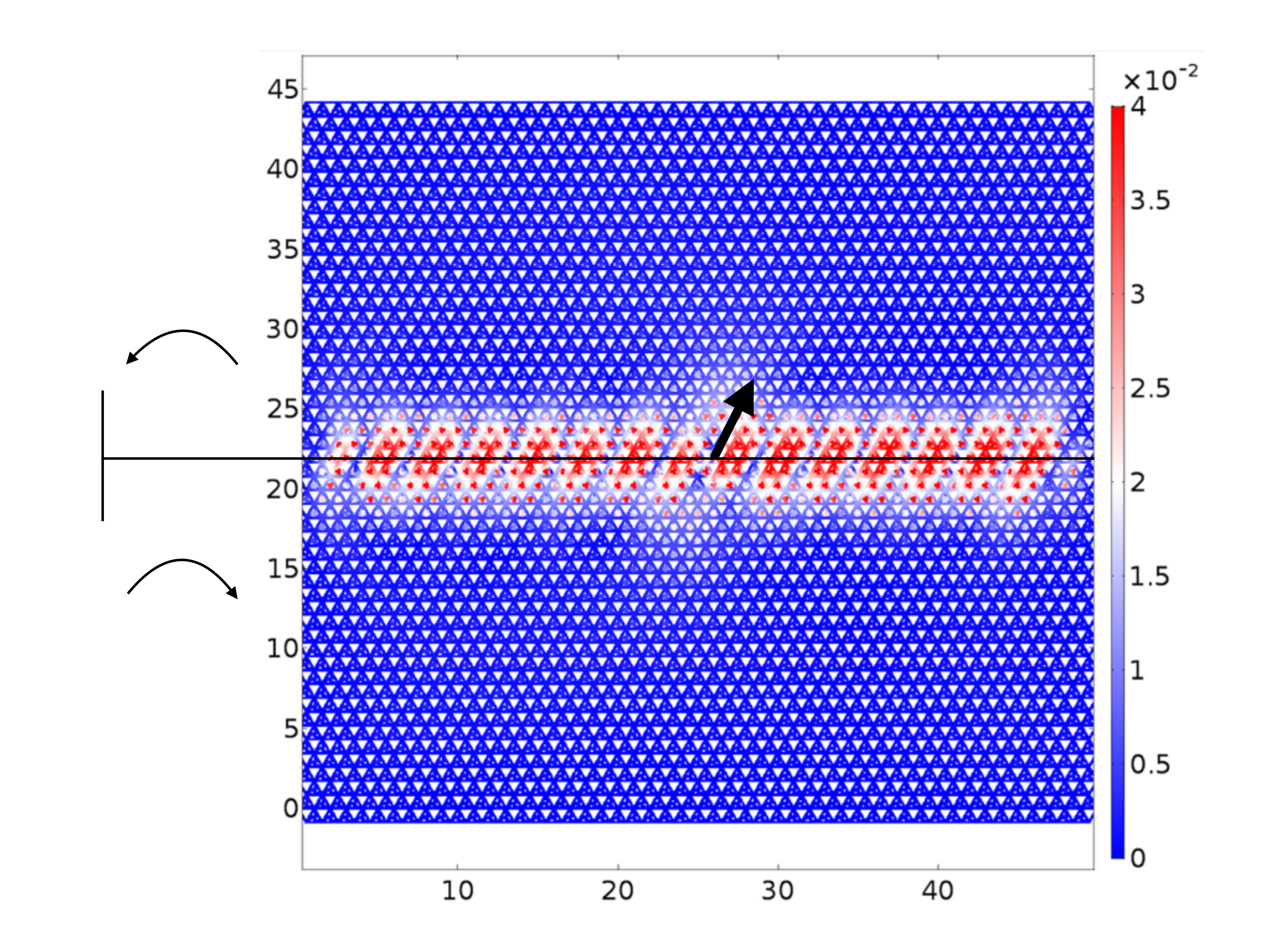}
	$$(d)\,\,\,\,\,\,\,\,\,\,\,\,\,\,\,\,\,\,\,\,\,\,\,\,\,\,\,\,\,\,\,\,\,\,\,\,\,\,\,\,\,\,\,\,\,\,\,\,\,\,\,\,\,\,\,\,\,\,\,\,\,\,\,\,\,\,\,\,\,\,\,\,\,\,\,\,\,\,\,\,(e)\,\,\,\,\,\,\,\,\,\,\,\,\,\,\,\,\,\,\,\,\,\,\,\,\,\,\,\,\,\,\,\,\,\,\,\,\,\,\,\,\,\,\,\,\,\,\,\,\,\,\,\,\,\,\,\,\,\,\,\,\,\,\,\,\,\,\,\,\,\,\,\,\,\,\,\,\,(f)$$
	\caption{\label{fig:point_load_DC} Panels (a), (b) and (c) are  the responses of a homogeneous TLR to a harmonic force of amplitude  $F=0.1~{\rm N}$ and frequency ${\omega}=\pi~{\rm rad/s}$ applied to a TL nodal point. The  point-loads (see black arrows) form an angle of $0$, $\pi/6$ and $\pi/3$, respectively, with respect to the horizontal axis.  Panels  (d), (e) and (f) are the responses of a non-homogeneous TLR  to harmonic forces identical to those considered in panels (a), (b) and (c), respectively.  The thin horizontal line marks the interface between anticlockwise tilting (upper part, $\vartheta_0=-0.82~{\rm rad}$) and clockwise tilting (lower part, $\vartheta_0=0.82~{\rm rad}$). The lattice parameters used in all panels - the same as in  Fig. (\ref{fig:disp_DC})(a) - are given in the first row of Table \ref{tab:parameters}.    }
\end{figure}
\begin{figure}[t!]
	\centering
	\includegraphics[width=0.5\textwidth]{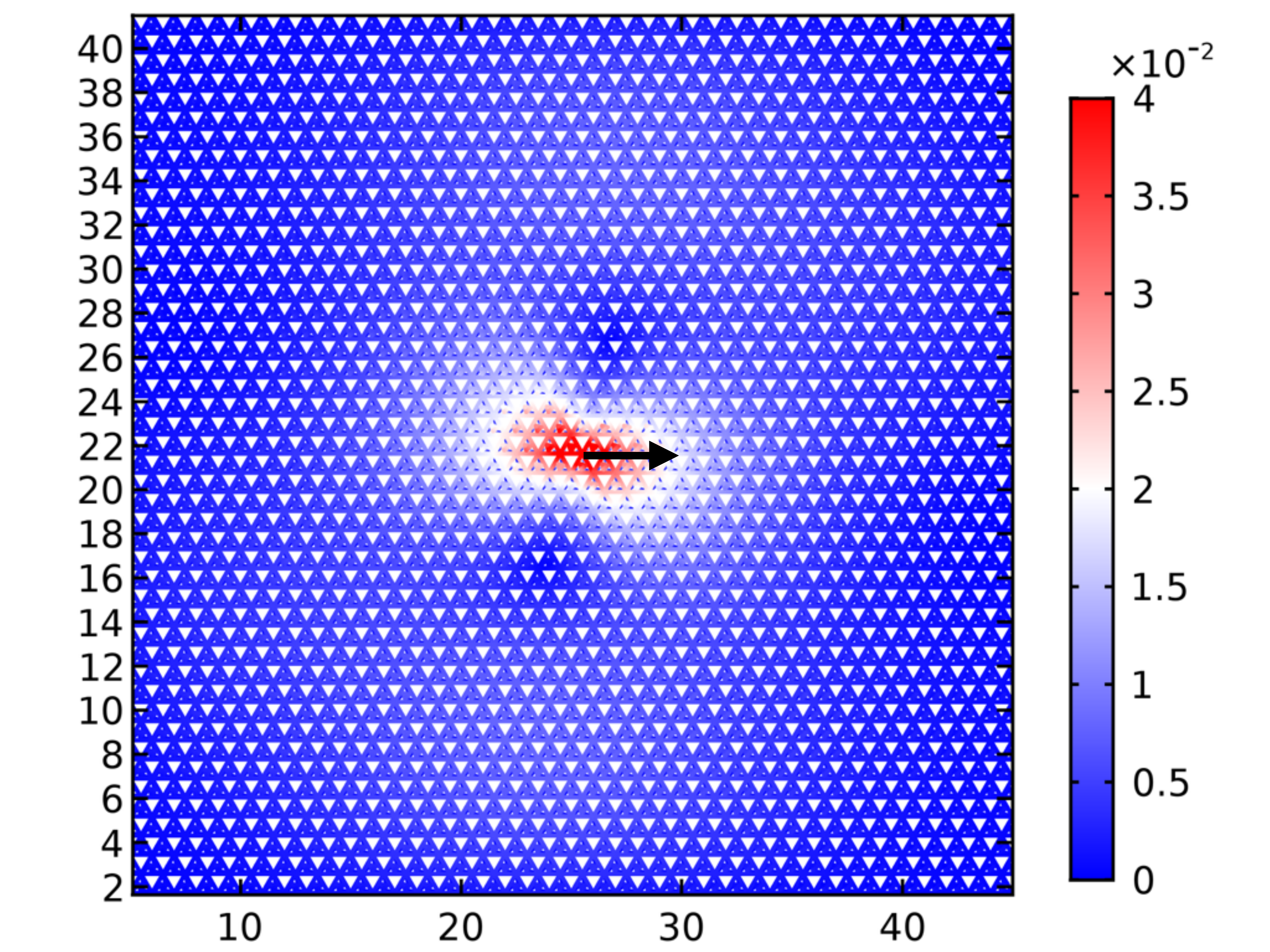}\hfill
	\includegraphics[width=0.5\textwidth]{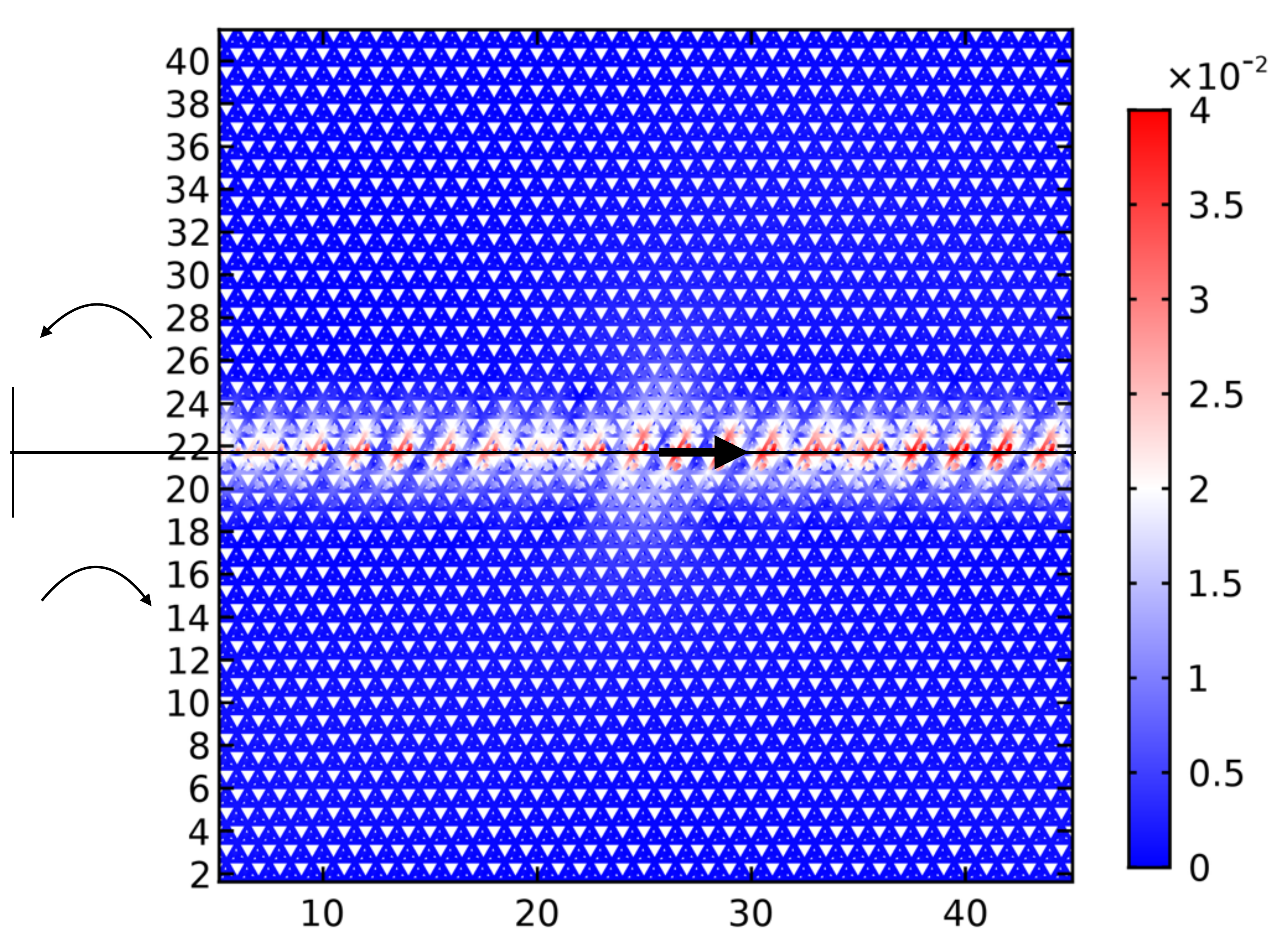}\\
	$(a)\,\,\,\,\,\,\,\,\,\,\,\,\,\,\,\,\,\,\,\,\,\,\,\,\,\,\,\,\,\,\,\,\,\,\,\,\,\,\,\,\,\,\,\,\,\,\,\,\,\,\,\,\,\,\,\,\,\,\,\,\,\,\,\,\,\,\,\,\,\,\,\,\,\,\,\,\,\,\,\,\,\,\,\,\,\,\,\,\,\,\,\,\,\,\,\,\,\,\,\,\,\,\,\,\,\,\,\,\,\,\,\,\,\,\,\,\,\,\,\,\,\,\,\,\,\,\,\,\,\,\,(b)$
	\includegraphics[width=0.5\textwidth]{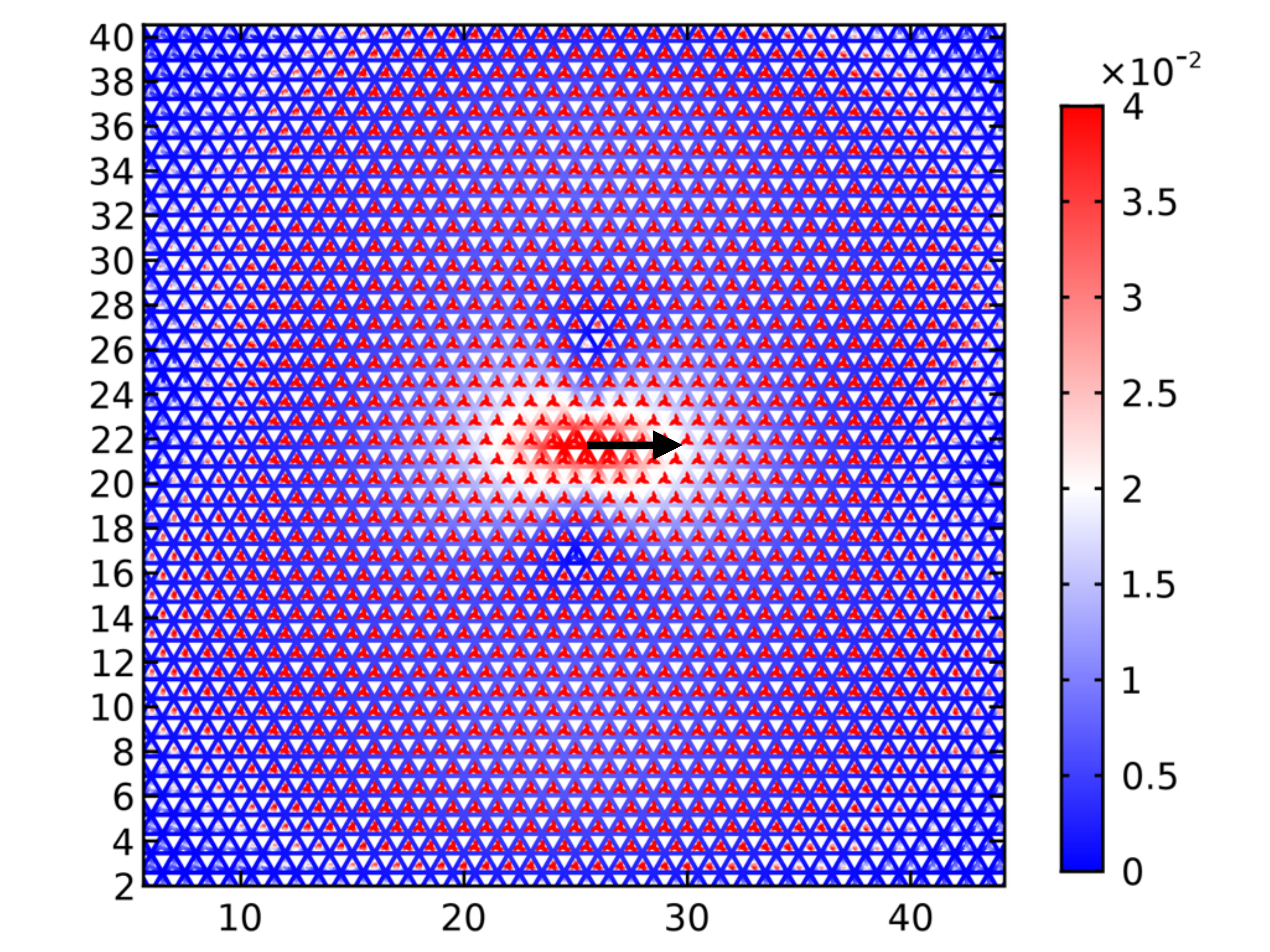}\hfill
	\includegraphics[width=0.5\textwidth]{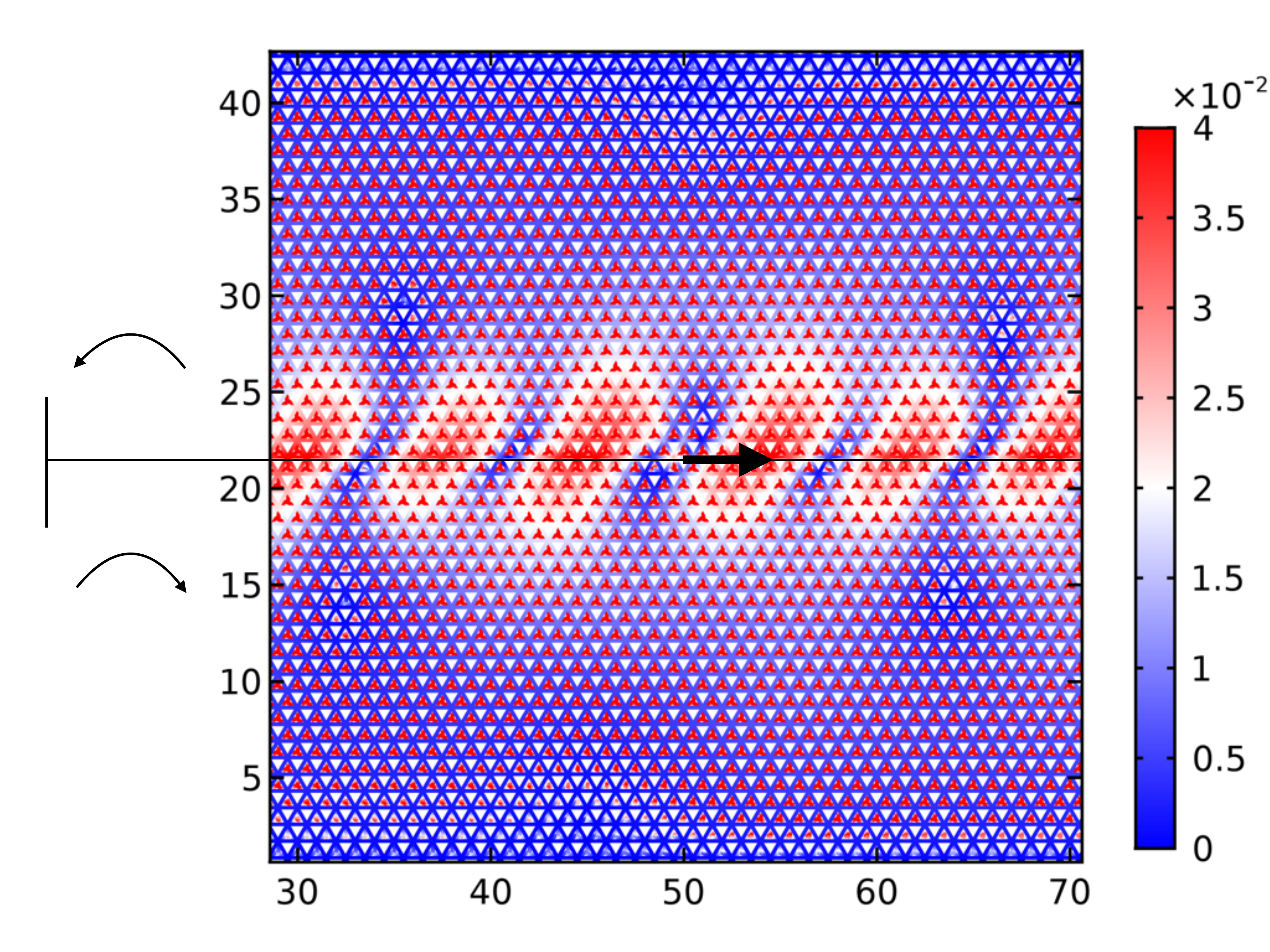}\\
	$(c)\,\,\,\,\,\,\,\,\,\,\,\,\,\,\,\,\,\,\,\,\,\,\,\,\,\,\,\,\,\,\,\,\,\,\,\,\,\,\,\,\,\,\,\,\,\,\,\,\,\,\,\,\,\,\,\,\,\,\,\,\,\,\,\,\,\,\,\,\,\,\,\,\,\,\,\,\,\,\,\,\,\,\,\,\,\,\,\,\,\,\,\,\,\,\,\,\,\,\,\,\,\,\,\,\,\,\,\,\,\,\,\,\,\,\,\,\,\,\,\,\,\,\,\,\,\,\,\,\,\,\,(d)$
	\caption{\label{fig:point_load_DC_comparison} Panels (a) and (c) are the responses of an homogeneous tilted cluster of resonators to an harmonic horizontal force of frequency $\omega=\pi~{\rm rad/s}$ and amplitude $F=0.1~{\rm N}$. The lattice parameters are the same as represented in Fig. \ref{fig:disp_DC}(d) by the red dashed line and the blue dotted line, respectively.  Panels (b) and (d) are the responses of a non-homogeneously tilted cluster of resonators to an harmonic horizontal force of frequency $\omega=\pi~{\rm rad/s}$ and amplitude $F=0.1~{\rm N}$. In panels (b) and (d), the tilting angle is oriented anticlockwise (clockwise) above (below) the thin horizontal line. The remaining lattice parameters, including the modulus of the tilting angle, are the same as in panel (a) and (c).}
\end{figure}

\section{Localisation and edge waves at the Dirac-like point \label{sec:EW_Loc}}
In this section,  we investigate the wave forms, which correspond to the frequencies in the neighbourhood of the Dirac-like point. In addition, we study  the propagation of edge waves along interfaces obtained by modifying the bulk homogeneous lattices. The periodic lattice's dynamic response to point loads of different orientations is studied using the Finite Element Method (COMSOL Multiphysics). In the computations we truncate the lattice retaining a $N\times N$ cluster of TLR's cells, where $N\approx 50$. \color{black} In order to reduce spurious reflections from the boundaries of the computational window, the dynamic equations of the nodal points close to the sides of the grid include a damping term. The damping layer has width $L_D=4L$  and is non uniform with spatial distribution $\eta(x)=\eta_0(1-{\rm exp}(-\sigma|x|))$, where $\sigma=1/L$ and $\eta_0$ is a frequency dependent factor and $x=[0,L_D]$ spans from the inner to the outer boundary of the damping frame. \color{black} The harmonic responses shown in this section are triggered by a point force of frequency  $\omega=\pi~{\rm rad/s}$, linear polarisation and amplitude $F=0.1~{\rm N}$. We assume that the force is exerted on a triangular lattice node, located at the centre of the clusters. The lattice parameters here considered are listed in Table \ref{tab:parameters}, where SI units of measurement and angles in unit of radiant are understood. These parameters have been chosen to reproduce a triple-eigenvalue at $\Gamma$ and frequency $\omega=\pi~{\rm rad/s}$  at the Dirac-like point (see section \ref{sec:DC}).

The effective properties of the dispersion surfaces emanating from the  Dirac-like point strongly influence the harmonic response of the structure. Special attention is given to the influence of the effective mass of the parabolic-in-${\bm k}$  mode, and to the effective group velocities of the  conical modes,  on the localisation patterns and on the amplitude and wavelength of the edge waves propagating along  interfaces obtained from the bulk TLRs. \color{black}

\subsection{Edge waves along the interface between non-homogeneously tilted TLR}

Figs \ref{fig:point_load_DC}(a), \ref{fig:point_load_DC}(b) and \ref{fig:point_load_DC}(c) show the harmonic responses of a cluster with lattice parameters as in set 1 of Table \ref{tab:parameters}. In these computations, three different linearly polarised forces have been used, each of which is oriented at $0$, $\pi/3$ and $\pi/6$ with respect to the horizontal axis (see black arrows). In Figs \ref{fig:point_load_DC}(a), \ref{fig:point_load_DC}(b) and \ref{fig:point_load_DC}(c), we observe a localisation pattern consistent with the flat band intersecting the Dirac cone at the triple eigenvalue.  The symmetry axis of the localisation pattern follows the polarisation angle of the force.  Figs \ref{fig:point_load_DC}(d), \ref{fig:point_load_DC}(e) and \ref{fig:point_load_DC}(f) show the harmonic responses of a special cluster of resonators in which an homogeneity has been introduced via the tilting angle. The remaining parameters are listed in ``set 1'' of Table \ref{tab:parameters} and the harmonic force is the same as in panels (a), (c) and (e). Above the thin black line the resonators are tilted in the anticlockwise direction ($\vartheta_0=-0.82$), while below the line a clockwise tilting ($\vartheta_0=0.82$) is implemented. This inhomogeneity introduces an interface which runs along the thin black line. It shall be pointed out that the dispersion surface of the lattice of resonators with clockwise and anticlockwise tilting are identical. In particular, the effective group velocities at the Dirac-like point are identical. 
Nevertheless, the harmonic response of the non-homogeneous cluster  differs significantly  from the corresponding responses of the homogeneously tilted cluster. In fact, we observe that a point force of frequency $\omega=\pi~{\rm rad/s}$, corresponding to the Dirac-like point, triggers an edge wave travelling along the interface. The amplitude of the edge wave depends on the orientation of the harmonic point force, being larger for larger deflections from the horizontal direction (\emph{cf.} Fig. \ref{fig:point_load_DC}(d) , \ref{fig:point_load_DC}(e) and \ref{fig:point_load_DC}(f)). The three panels suggest that the elastic edge wave propagating along the interface have elliptic polarisation whose principal axis is oriented at  $\pi/3$ with respect to the interface. When the linear polarisation angle of the source matches $\pi/3$ (see Fig. \ref{fig:point_load_DC}(f)), the amplitude of the edge wave is  greater than the other two cases for geometrical reasons.
\color{black}

In the same spirit as in Figs \ref{fig:point_load_DC}, Figs  \ref{fig:point_load_DC_comparison} show the harmonic responses of clusters whose lattice parameters are listed in  ``set 2'' (panels (a) and (b)) and  ``set 3'' (panels (c) and (d)) of Table \ref{tab:parameters}. The aim  here is to illustrate how different dispersive properties near the Dirac-like point, already highlighted in Fig. \ref{fig:disp_DC}(d),  affect the harmonic responses of homogeneously tilted clusters (panels (a) and (c)) and non-homogeneously tilted clusters (panels (b) and (d)). The  non-homogeneity considered here has the same meaning as in Figs \ref{fig:point_load_DC}.  Figs \ref{fig:point_load_DC_comparison}(a) and \ref{fig:point_load_DC_comparison}(c) show localised patterns similar to what encountered in Fig. \ref{fig:point_load_DC}(a). Figs \ref{fig:point_load_DC_comparison}(c) and \ref{fig:point_load_DC_comparison}(d) show an edge wave travelling across the interface. We remark that the wavelength of the edge waves is larger for smaller effective group velocities at the Dirac-like point $\omega=\pi~{\rm rad/s}$. This suggests that the dynamics of the edge waves is controlled by the effective group velocities  at the Dirac-like point.
\begin{figure}[h!]
	\centering
	\includegraphics[width=0.8\textwidth]{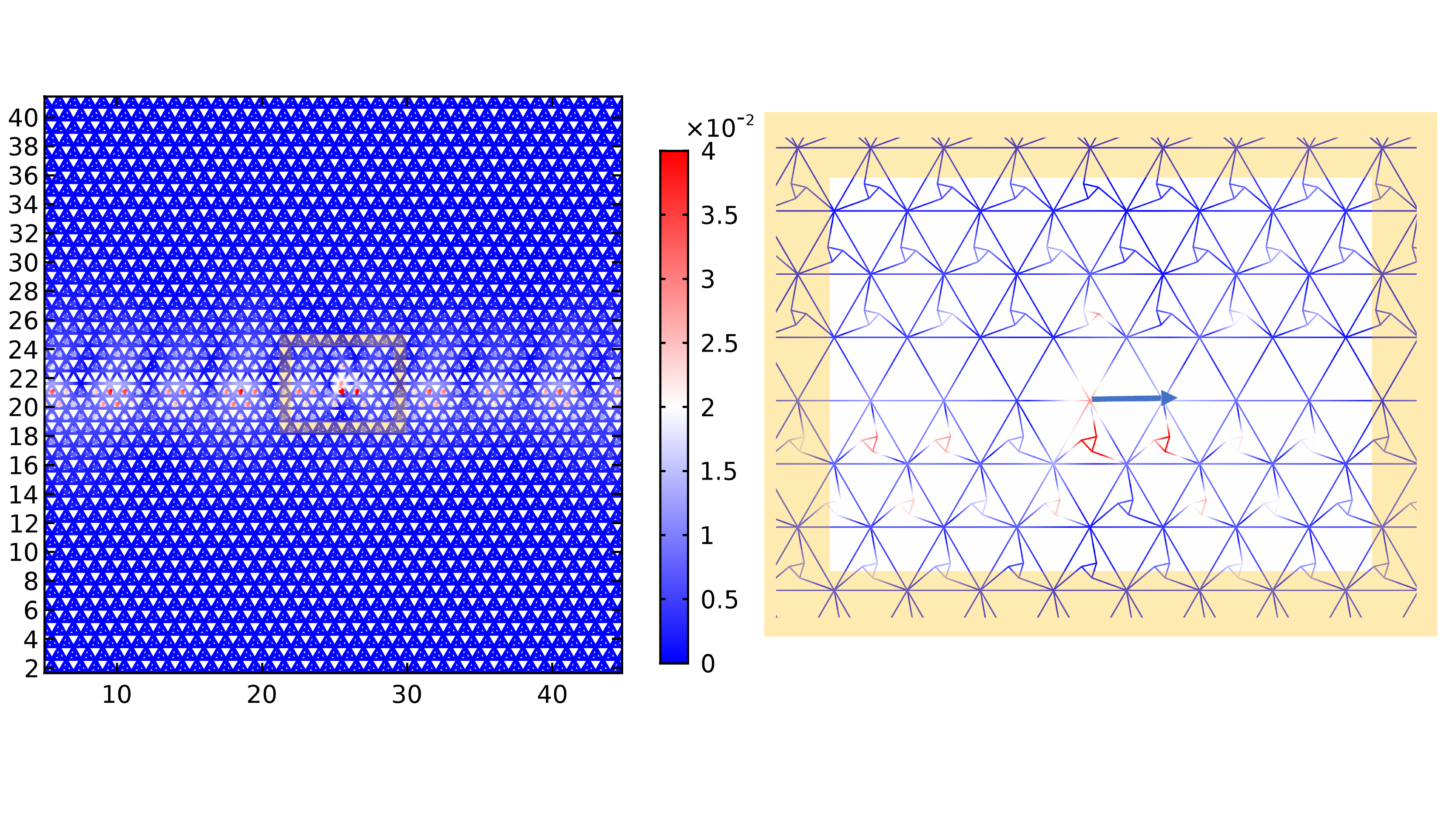}
	\caption{\label{fig:point_load_DC_line} Response of a defective triangular lattice with resonators to an harmonic point force. The defect consists of line along which resonators are removed, as highlighted in the yellow magnifying inset on the right.  The point force is represented in the right inset by the blue arrow and has amplitude  $F=0.1~{\rm N}$ and frequency $\omega=\pi~{\rm rad/s}$. The parameters used in this computation are listed in the first row of Table \ref{tab:parameters}  and the tilting angle is anticlockwise and clockwise, above and below the defect, respectively.\color{black}}
\end{figure}
\subsection{Edge waves along a line defect in a non-homogeneously tilted TLR}
Fig. \ref{fig:point_load_DC_line} shows the modulus of the displacement field for a forced TLR contaning a  defect which consists of a missing line of resonators, as shown in the magnifying inset highlighted in yellow on the left of the figure. The lattice parameters used in this computation are listed in set 1 of Table \ref{tab:parameters} and the tilting angle is anticlockwise and clockwise, above and below the defect, respectively. The harmonic force is identical to the one used in Fig. \ref{fig:point_load_DC}(a) and is exerted on a triangular lattice nodal point below the line defect (see blue arrow in the inset). We observe that the defect acts as a wave guide for an edge wave whose wavelength differs from the one in Fig. \ref{fig:point_load_DC}(b). We emphasise again that the wave-guiding behaviour in Fig. \ref{fig:point_load_DC_line} differs significantly from the localisation pattern in Fig. \ref{fig:point_load_DC}(a), the bulk homogeneous counterpart.

\begin{figure}[h!]
	\centering
		\includegraphics[trim={5.5cm 12.5cm  5.5cm 11.5cm},width=0.5\textwidth]{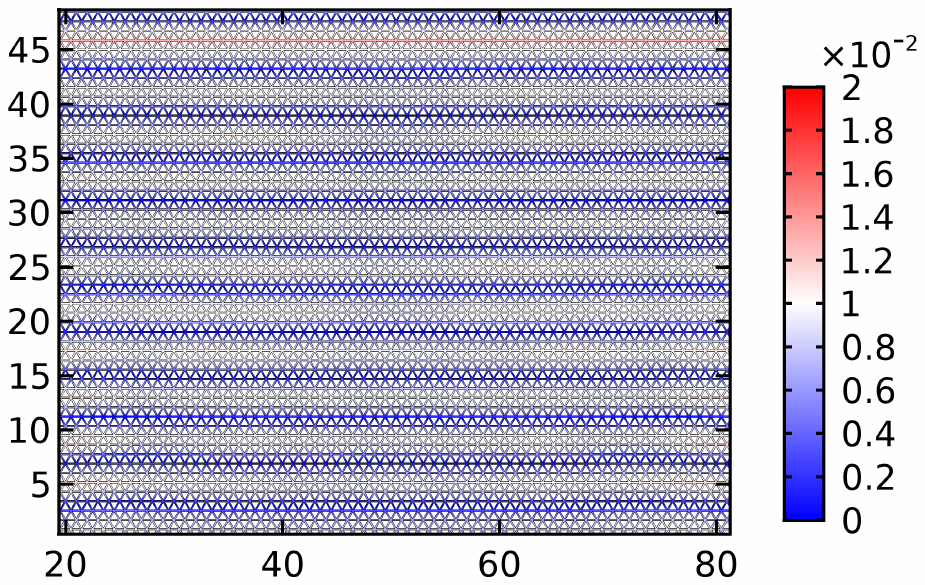}\hfill
		\includegraphics[trim={5.5cm 12.5cm  5.5cm 11.5cm},width=0.5\textwidth]{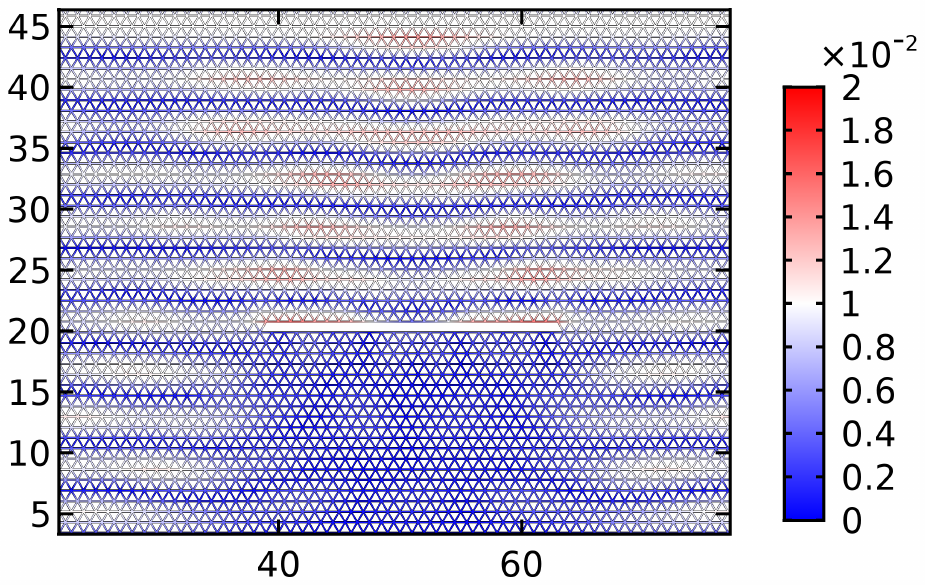}\\
	$$(c)\,\,\,\,\,\,\,\,\,\,\,\,\,\,\,\,\,\,\,\,\,\,\,\,\,\,\,\,\,\,\,\,\,\,\,\,\,\,\,\,\,\,\,\,\,\,\,\,\,\,\,\,\,\,\,\,\,\,\,\,\,\,\,\,\,\,\,\,\,\,\,\,\,\,\,\,\,\,\,\,\,\,\,\,\,\,\,\,\,\,\,\,\,\,\,\,\,\,\,\,\,\,\,\,\,\,\,\,\,\,\,\,\,\,\,\,\,\,\,\,\,\,\,\,\,\,\,\,\,\,\,\,\,\,\,\,\,\,\,\,\,\,\,\,\,\,\,\,\,\,\,\,(d)$$
	\caption{\label{fig:PW+uncoated_crack} Panel (a) shows a shear plane wave of angular frequency $\omega=\pi~{\rm rad/s}$  travelling through an homogeneous triangular lattice. In panel (b) the same shear wave is scattered by a one dimensional uncoated crack.}
\end{figure}
\section{Wave-forms around a crack surrounded by a micro-structured coating \label{sec:coated_crack} }  
In this section we study a special coating for one-dimensional cracks inside a TL. We consider a shear plane wave of angular frequency $\omega=\pi~{\rm rad/s}$ impinging on the crack. The coating is obtained by introducing resonators  around the crack. 

The physical parameters of the exterior triangular lattice in which the plane wave propagates,  can be chosen in order to guarantee an isotropic dynamic response. In this section, the maximum plane wave' s frequency is $\omega=\pi~{\rm rad/s}$. The stiffness of the links $c_{\rm TL}=50~{\rm N/m}$, for the mass of the nodal points corresponding to $m_{\rm TL}=m+3m_o=1.43~{\rm Kg}$ (see set 1 in Table \ref{tab:parameters}),  guarantees an  isotropic dynamic response.  We observe that the aforementioned choice of the mass minimises spurious scattering effects associated with a contrast of inertia. Fig. \ref{fig:PW+uncoated_crack}(a) shows a shear plane wave of frequency $\omega=\pi~{\rm rad/s}$ propagating through the isotropic triangular lattice.  The wave is excited by applying a time-harmonic horizontal displacement to the nodal points of the lattice close to the horizontal line $y=45$. In Fig. \ref{fig:PW+uncoated_crack}(b), a crack obtained by removing some links from the triangular lattice scatters the shear plane wave. 

In this section, the lattice parameters of  the structured coating  are given in set 1 of Table \ref{tab:parameters}. The corresponding dispersion surfaces are reported in Fig. \ref{fig:disp_DC}(a). The different frequency regimes are discussed via the analysis of the scattered displacement fields: in subsection \ref{subsec:dc} we address frequencies close to the Dirac-like point and  in subsection \ref{subsec:bg} we focus on band gap regime.

\begin{figure}[h!]
	\centering
	\includegraphics[trim={5.5cm 11.5cm  5.5cm 11.5cm},width=0.5\textwidth]{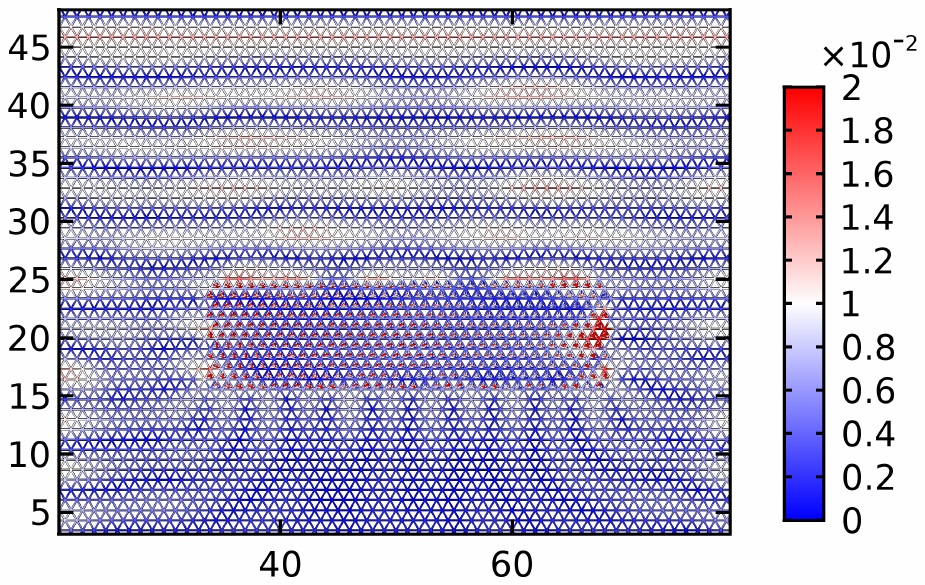}\hfill
	\includegraphics[trim={5.5cm 11.5cm  5.5cm 11.5cm},width=0.5\textwidth]{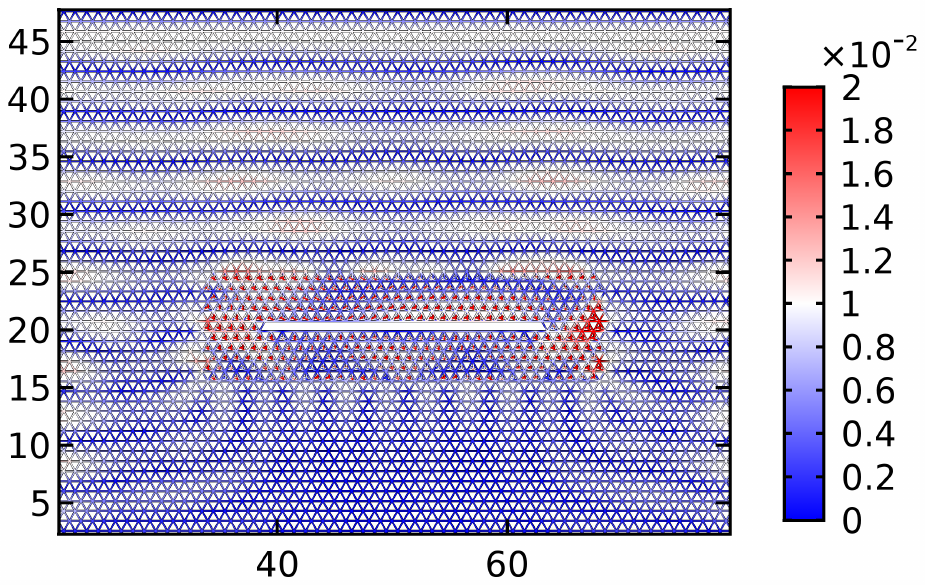}
	$(a)\,\,\,\,\,\,\,\,\,\,\,\,\,\,\,\,\,\,\,\,\,\,\,\,\,\,\,\,\,\,\,\,\,\,\,\,\,\,\,\,\,\,\,\,\,\,\,\,\,\,\,\,\,\,\,\,\,\,\,\,\,\,\,\,\,\,\,\,\,\,\,\,\,\,\,\,\,\,\,\,\,\,\,\,\,\,\,\,\,\,\,\,\,\,\,\,\,\,\,\,\,\,\,\,\,\,\,\,\,\,\,\,\,\,\,\,\,\,\,\,\,\,\,\,\,\,\,\,\,\,\,\,\,\,\,\,\,\,\,(b)$
	\caption{\label{fig:DC_T} The harmonic responses to a shear plane wave of frequency ${\omega}=\pi~{\rm rad/s}$ corresponding to the Dirac-like point for the TLR. Panels (a) and (b) represent a cluster of resonators and a crack surrounded by a cluster of resonators, respectively.  The parameters used to model the clusters are listed in set 1 of Table \ref{tab:parameters}.}
\end{figure}
\begin{figure}[h!]
	\centering
	\includegraphics[trim={5.5cm 13.5cm  5.5cm 13cm},width=0.5\textwidth]{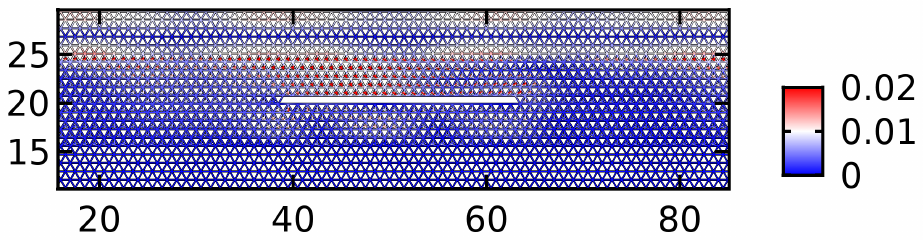}\hfill
	\includegraphics[trim={5.5cm 13.5cm  5.5cm 13cm},width=0.5\textwidth]{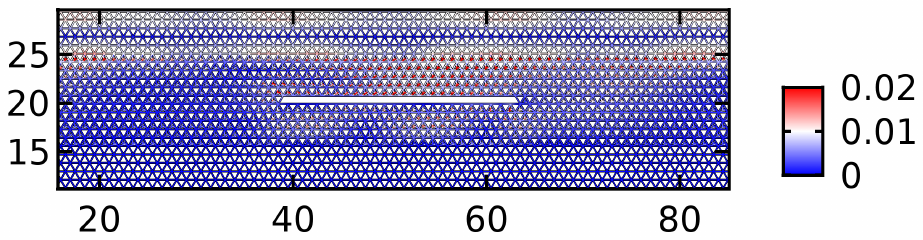}
$(a)\,\,\,\,\,\,\,\,\,\,\,\,\,\,\,\,\,\,\,\,\,\,\,\,\,\,\,\,\,\,\,\,\,\,\,\,\,\,\,\,\,\,\,\,\,\,\,\,\,\,\,\,\,\,\,\,\,\,\,\,\,\,\,\,\,\,\,\,\,\,\,\,\,\,\,\,\,\,\,\,\,\,\,\,\,\,\,\,\,\,\,\,\,\,\,\,\,\,\,\,\,\,\,\,\,\,\,\,\,\,\,\,\,\,\,\,\,\,\,\,\,\,\,\,\,\,\,(b)$
\vfill
\includegraphics[trim={5.5cm 13.5cm  5.5cm 13cm},width=0.5\textwidth]{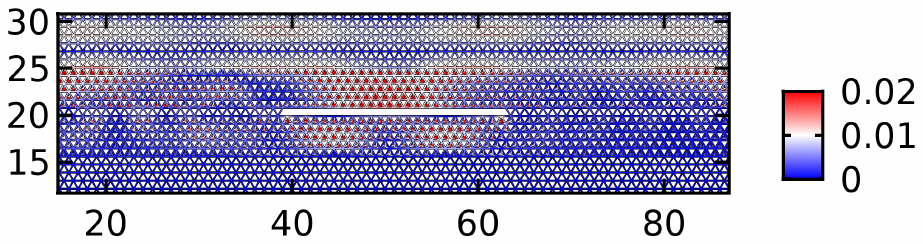}\hfill
\includegraphics[trim={5.5cm 13.5cm  5.5cm 13cm},width=0.5\textwidth]{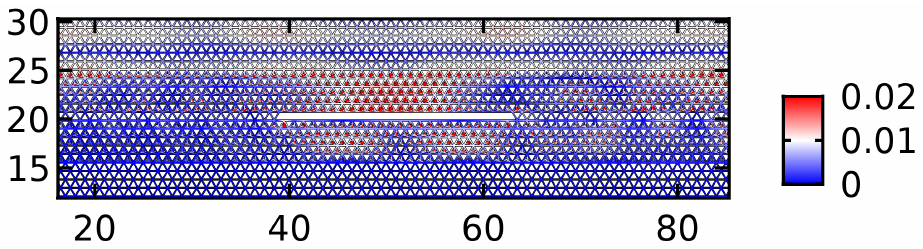}
$(c)\,\,\,\,\,\,\,\,\,\,\,\,\,\,\,\,\,\,\,\,\,\,\,\,\,\,\,\,\,\,\,\,\,\,\,\,\,\,\,\,\,\,\,\,\,\,\,\,\,\,\,\,\,\,\,\,\,\,\,\,\,\,\,\,\,\,\,\,\,\,\,\,\,\,\,\,\,\,\,\,\,\,\,\,\,\,\,\,\,\,\,\,\,\,\,\,\,\,\,\,\,\,\,\,\,\,\,\,\,\,\,\,\,\,\,\,\,\,\,\,\,\,\,\,\,\,\,(d)$
	\caption{\label{fig:DC_long} 
		\color{black} The harmonic responses to a shear plane wave of frequency ${\omega}=\pi~{\rm rad/s}$ corresponding to the Dirac point for the triangular lattice with resonators. In panels (a) and (b), we substitute the cluster of Fig. \ref{fig:DC_T}(b), which is finite in the horizontal direction,  with an infinite  strip. In panels (a) and (b), the tilting is clockwise and anticlockwise, respectively.  The structure in panels (c) and (d) is obtained from panel (a) by removing a horizontal line of resonators along the extension of the crack. In panel (c) an homogeneous tilting is used; in panel (d) the resonators above the line are rotated anticlockwise and those below clockwise.}
\end{figure}
\subsection{Dirac-like regime \label{subsec:dc}} 

In Fig. \ref{fig:DC_T} we compare the modulus of the displacement field resulting from the interaction of an elastic shear wave with  a cluster of resonators (panel (a)) and with a cluster of resonators containing a crack (panel (b)). The source of the excitation is a plane wave of frequency $\omega=\pi~{\rm rad/s}$ which corresponds to the Dirac-like point for the periodic TLR (see Fig. \ref{fig:disp_DC}). Panel (a) shows that scattering of elastic waves is highly anisotropic, the displacement field being concentrated on the right side  of the cluster. It is worthwhile noting that if the resonators are rotated in the anticlockwise direction ($\vartheta_0=-0.82$) the displacement field is  mirror-symmetric compared to the one of Fig. \ref{fig:DC_T}(a). The introduction of a crack within the cluster (Fig. \ref{fig:DC_T}(b)) triggers the propagation of elastic waves around the crack itself. The displacement field and the corresponding stresses are still visibly concentrated around the right tip of the crack. This suggests that a coating of resonators in the Dirac-like regime is likely to lead to a  left-right asymmetry in the propagation of the crack. 

In Figs \ref{fig:DC_long},  long strips of resonators containing a crack interact with a shear plane wave impinging on the strip from above. Several arrangements for the resonators are considered. In panel (a) (panel (b)) the resonators in the strip are homogeneously tilted in the clockwise (anticlockwise) direction. Similarly to Fig. \ref{fig:DC_T}(a), this leads to an enhancement of the displacement field close to the tips of the cracks. Moreover, the results are  mirror-symmetric about the vertical line passing through the centre of the crack. This is consistent with what we observe in Fig. \ref{fig:DC_T}(b). In Fig. \ref{fig:DC_long}(c) the homogeneously tilted strip analysed in panel (a) has been replaced by a strip with an interface. The interface is represented by a line of  missing resonators. The stiffness of the triangular lattice links which define the interface, is assumed to be $c_{\rm TL}=50~{\rm N/m}$, as in the exterior triangular lattice. In Fig. \ref{fig:DC_long}(d), the strip is similar to the one of Fig. \ref{fig:DC_long}(c) but  anticlockwise tilting above the line and clockwise tilting below the line is implemented. In Figs \ref{fig:DC_long}(c) and \ref{fig:DC_long}(d) the displacement field is mirror-symmetric with respect to a vertical line passing through  the crack. 
\begin{figure}[h!]
	\centering
	\includegraphics[trim={3.5cm 10cm  5.5cm 8.5cm},width=0.48\textwidth]{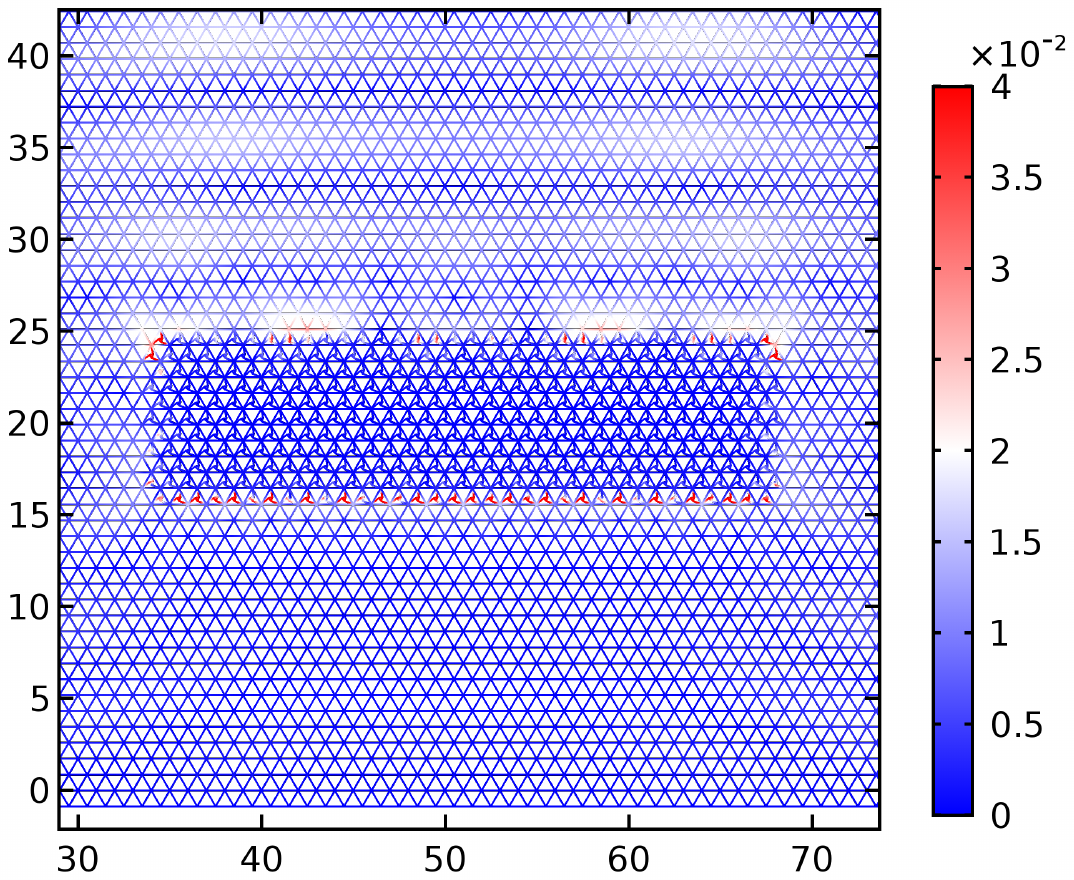}\hfill
	\includegraphics[trim={3.5cm 10cm  5.5cm 8.5cm},width=0.48\textwidth]{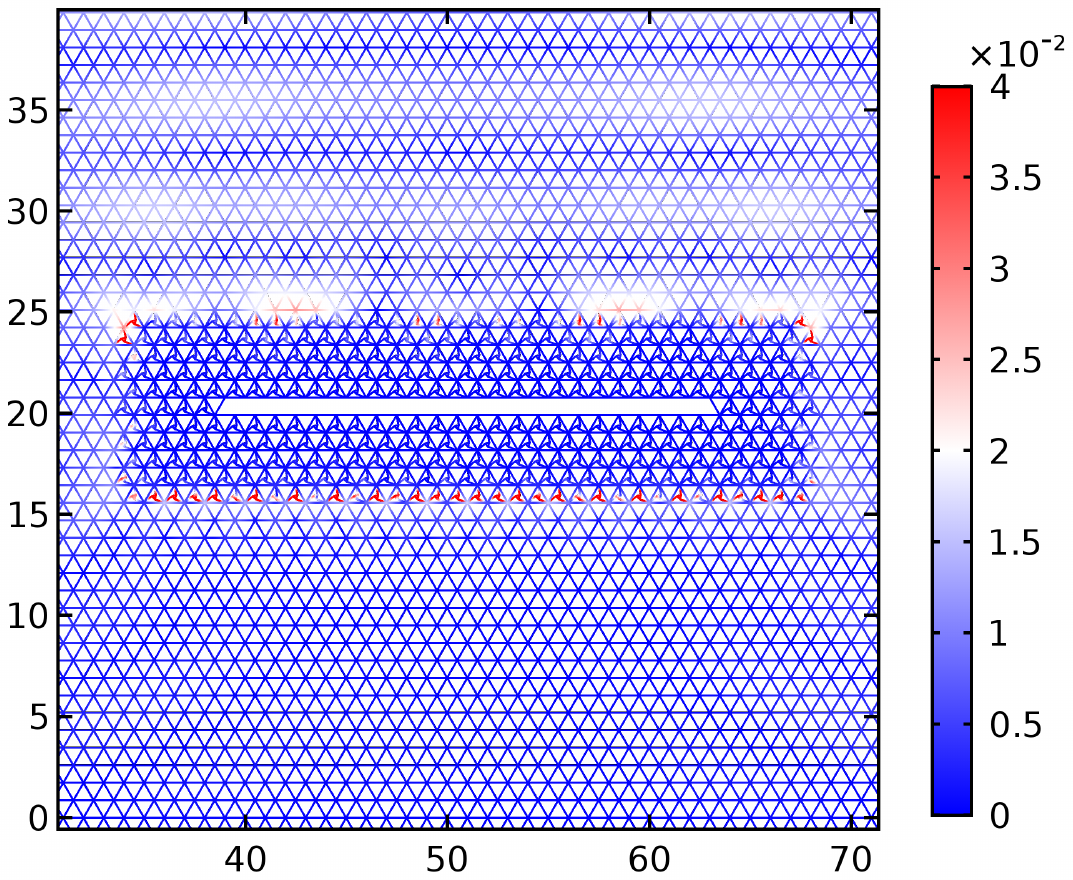}\\
	$(a)\,\,\,\,\,\,\,\,\,\,\,\,\,\,\,\,\,\,\,\,\,\,\,\,\,\,\,\,\,\,\,\,\,\,\,\,\,\,\,\,\,\,\,\,\,\,\,\,\,\,\,\,\,\,\,\,\,\,\,\,\,\,\,\,\,\,\,\,\,\,\,\,\,\,\,\,\,\,\,\,\,\,\,\,\,\,\,\,\,\,\,\,\,\,\,\,\,\,\,\,\,\,\,\,\,\,\,\,\,\,\,\,\,\,\,\,\,\,\,\,\,\,\,\,\,\,\,\,\,\,\,\,(b)$
	\\
		\includegraphics[trim={0 0  7cm 1cm},width=0.2\textwidth]{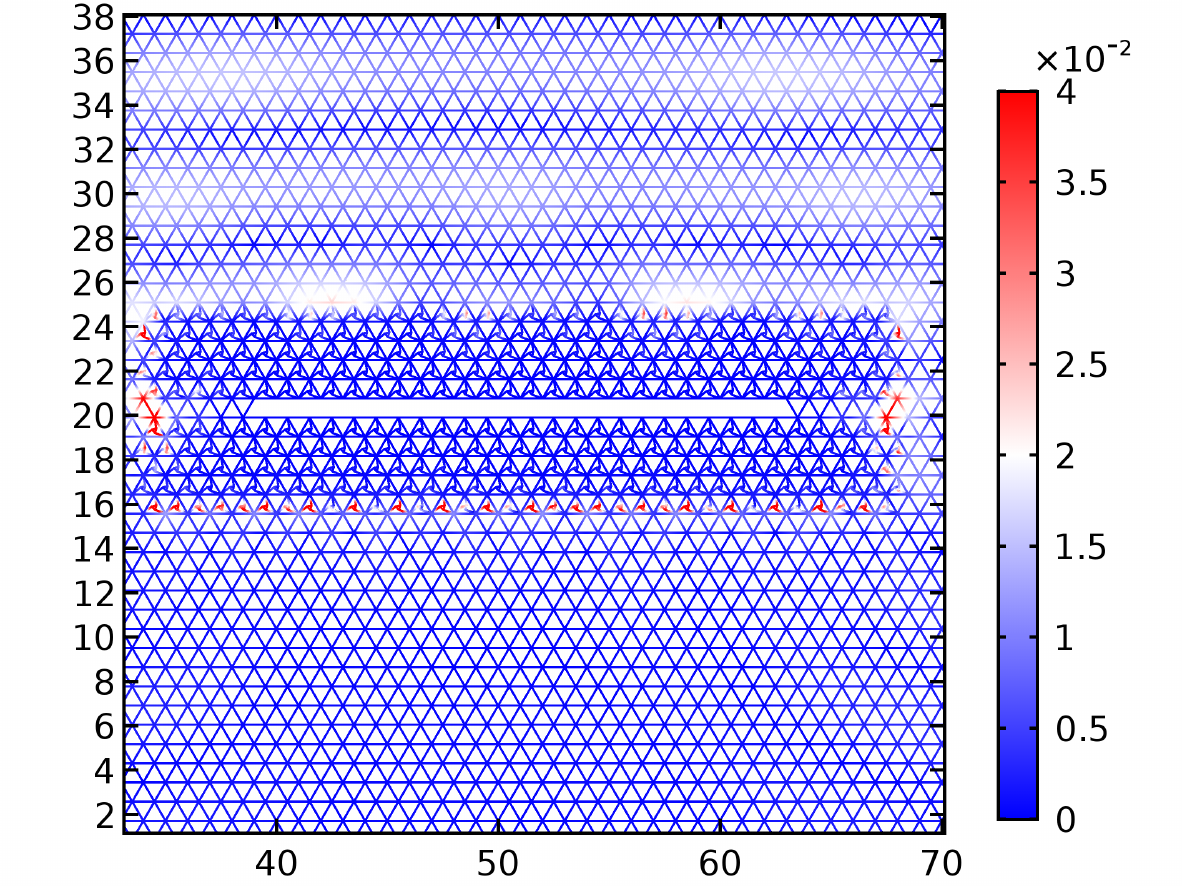}\hfill
	\includegraphics[trim={3.5cm 10.3cm  5.5cm 9.5cm},width=0.48\textwidth]{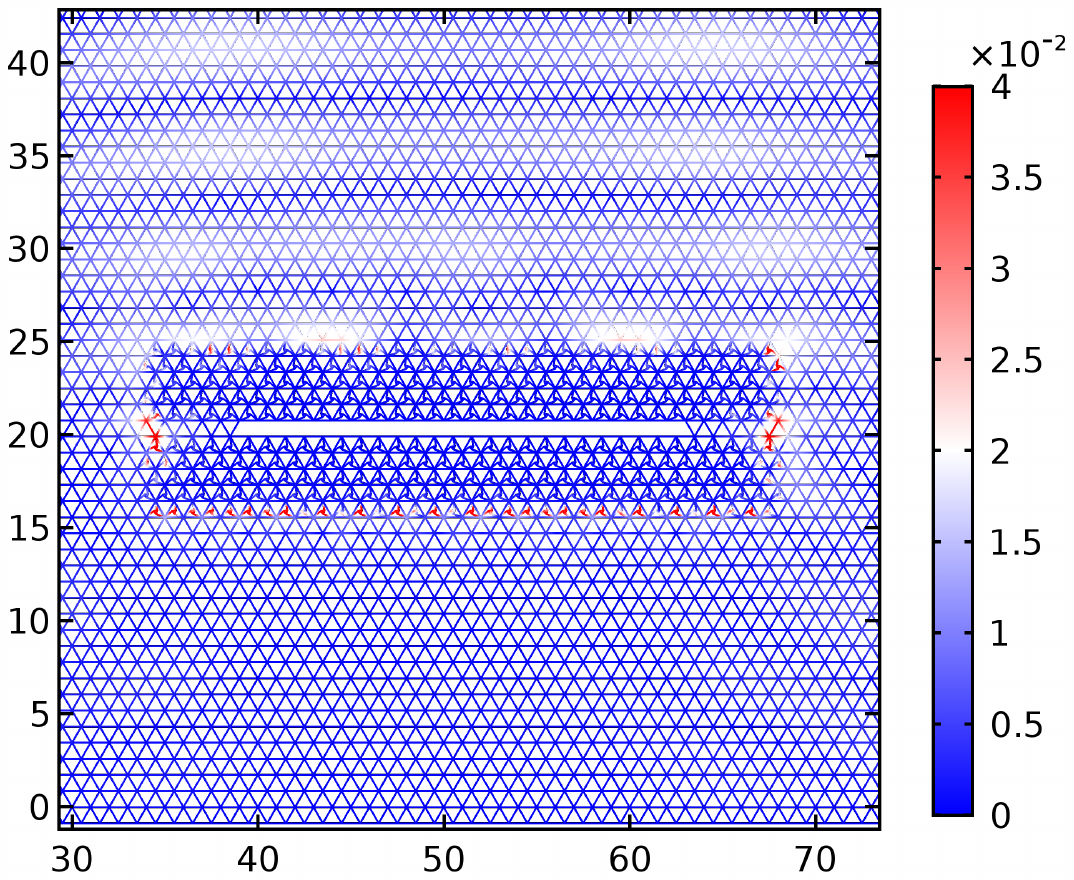}\\
$(c)\,\,\,\,\,\,\,\,\,\,\,\,\,\,\,\,\,\,\,\,\,\,\,\,\,\,\,\,\,\,\,\,\,\,\,\,\,\,\,\,\,\,\,\,\,\,\,\,\,\,\,\,\,\,\,\,\,\,\,\,\,\,\,\,\,\,\,\,\,\,\,\,\,\,\,\,\,\,\,\,\,\,\,\,\,\,\,\,\,\,\,\,\,\,\,\,\,\,\,\,\,\,\,\,\,\,\,\,\,\,\,\,\,\,\,\,\,\,\,\,\,\,\,\,\,\,\,\,\,\,\,\,(d)$\\
	\includegraphics[trim={8cm 5cm  8cm 5cm},width=0.15\textwidth,angle =270]{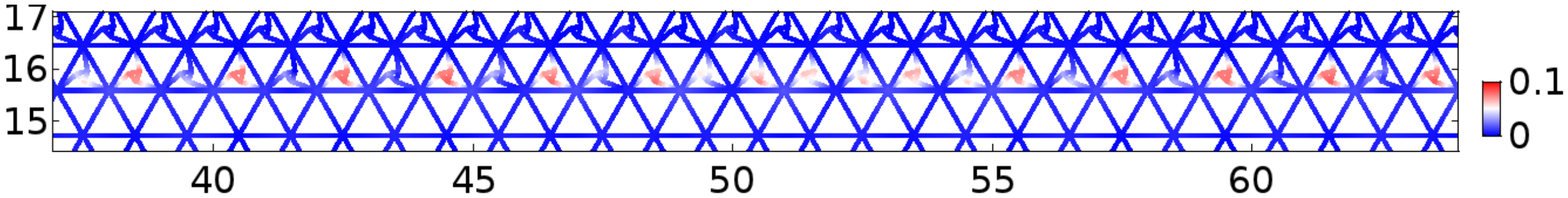}\\
	$(e)$
	\caption{\label{fig:PW_T_Gap} The harmonic response of a cluster of resonators to a shear plane wave of frequency ${\omega}=2.4~{\rm rad/s}$ inside the stop band for the TLR. Panel (a) and (b) are  without and with a crack. Panels (c) and (d) include a line of resonators missing along the extension of the crack. In panel (c) the tilting angle is homogeneous,  whereas in panel (d) the resonators are tilted through opposite angles. Panel (e) is a detail of the lower boundary of the cluster in panel (a) showing an  edge wave. The lattice parameters are as in Figs \ref{fig:DC_T}.}
\end{figure}
\begin{figure}[h!]
	\centering
	\includegraphics[width=0.23\textwidth,angle=270]{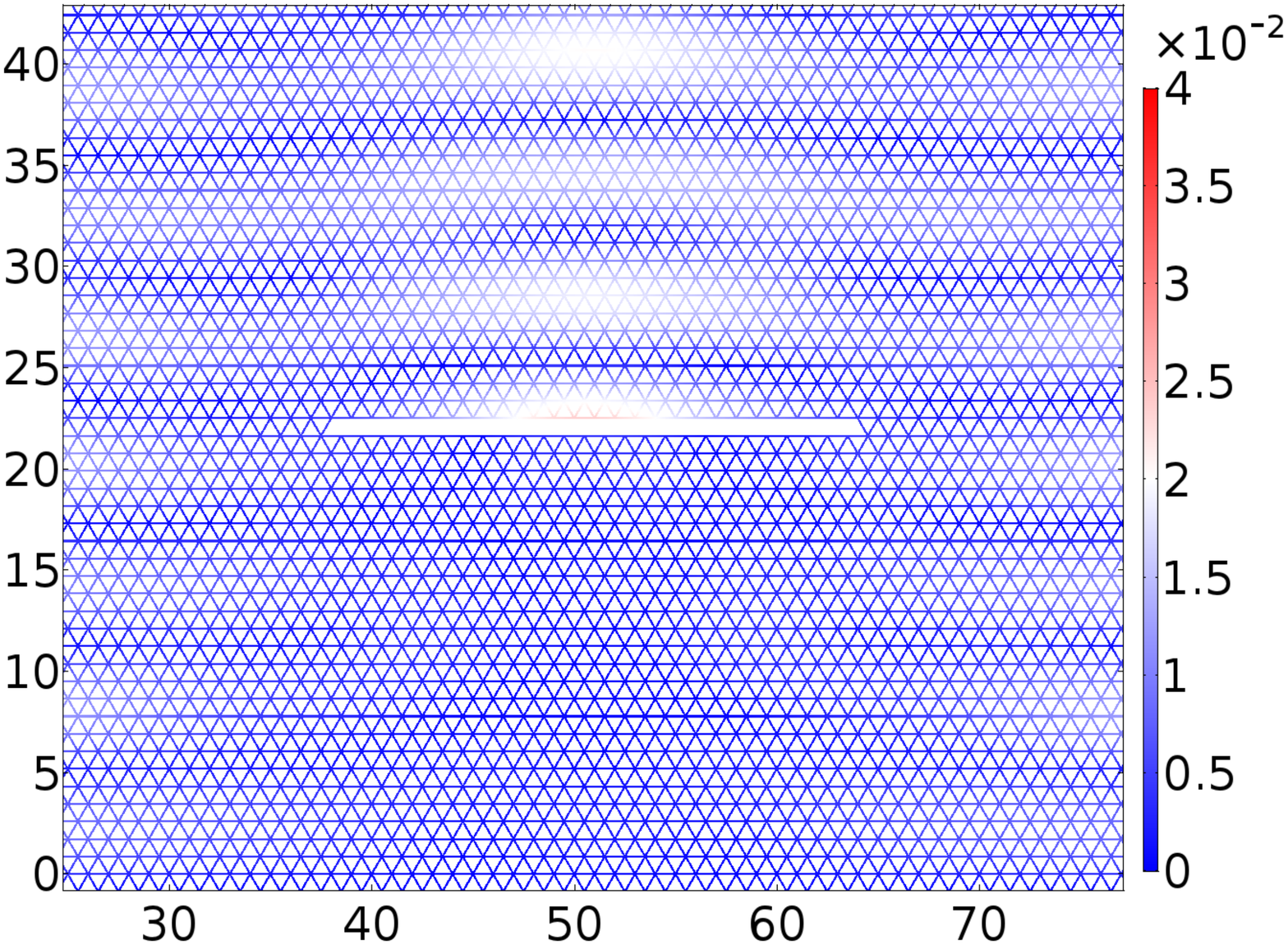}\hfill
	\includegraphics[width=0.23\textwidth,angle=270]{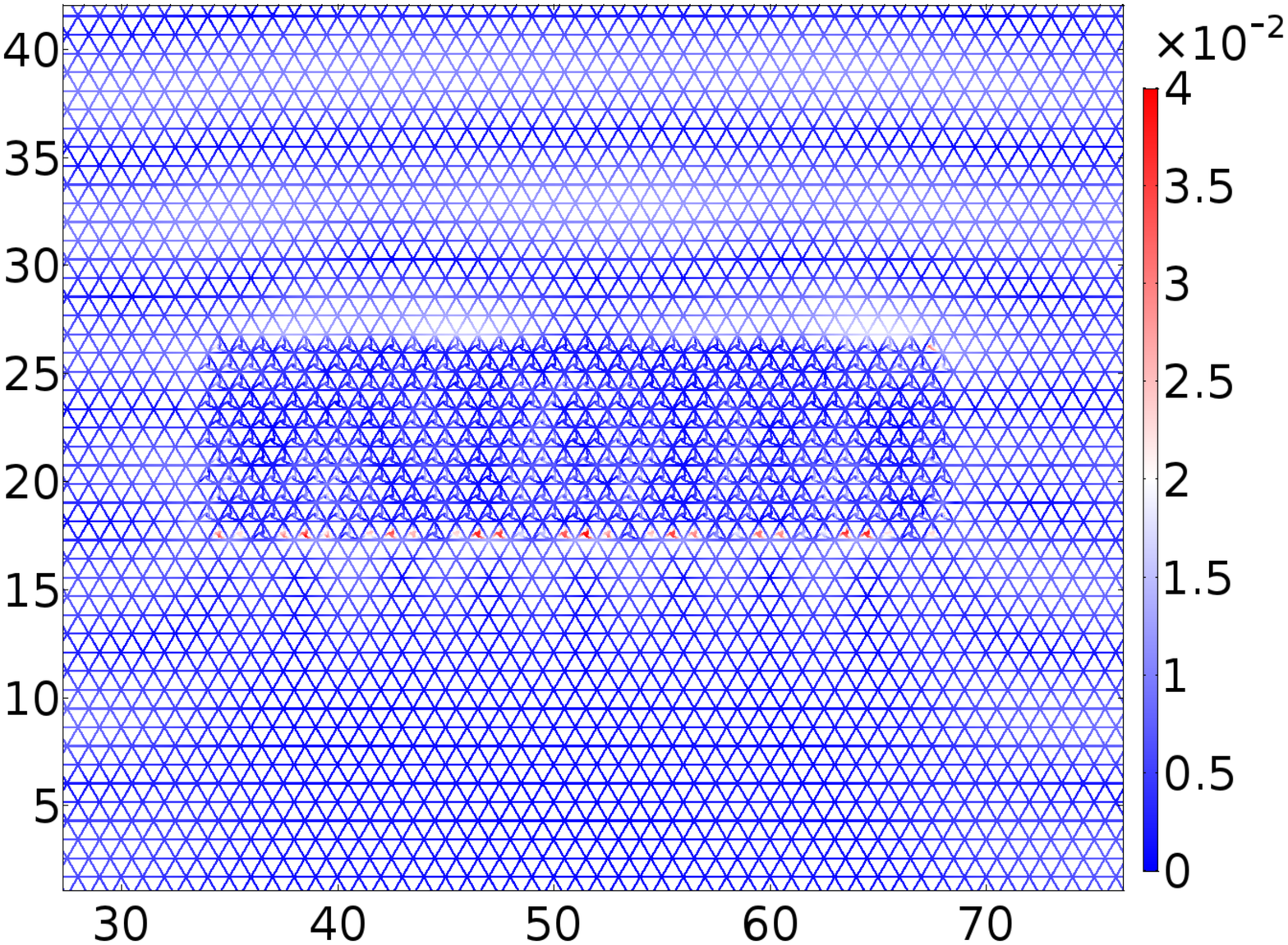}\hfill
	\includegraphics[width=0.23\textwidth,angle=270]{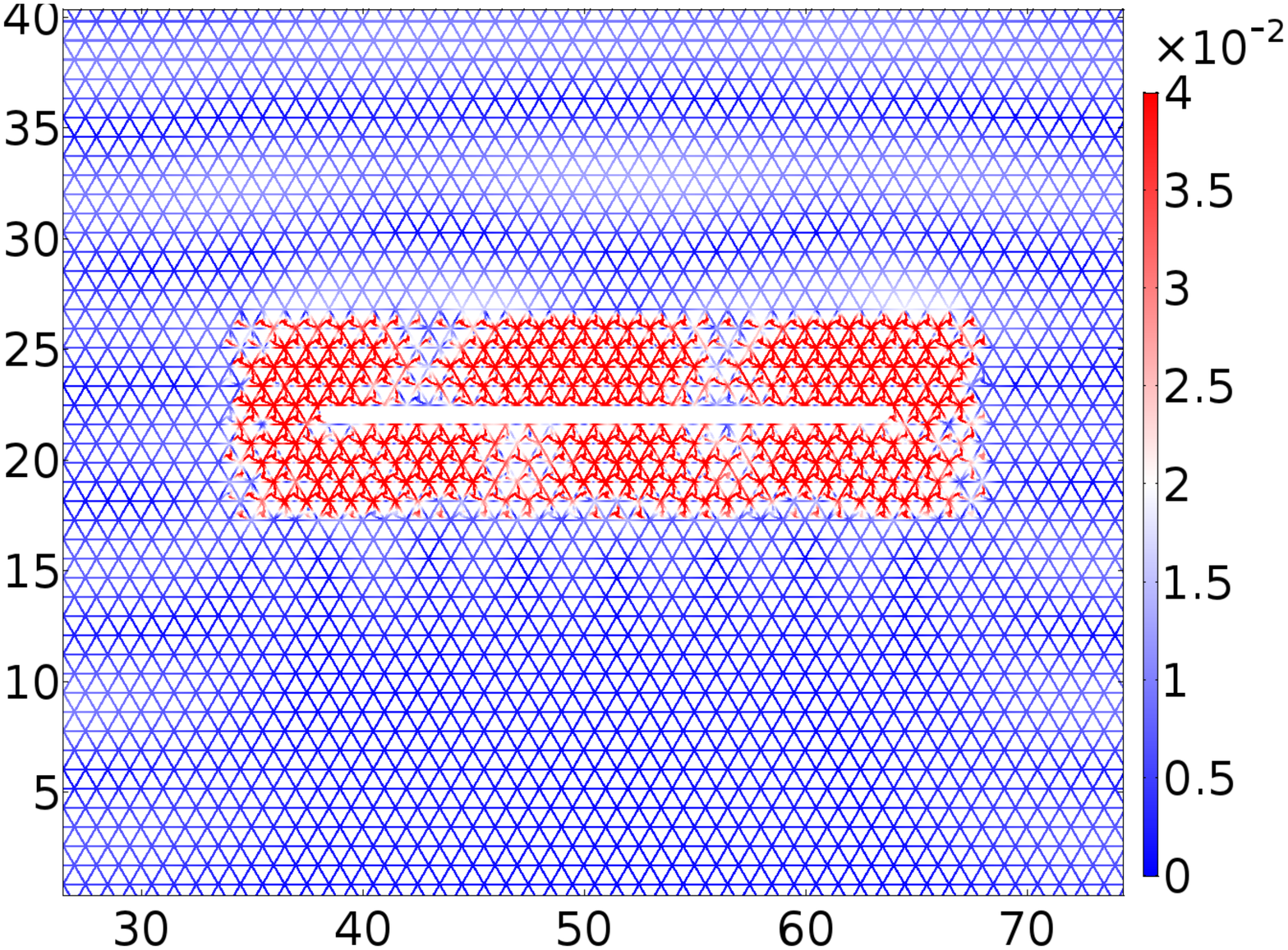}\\
	$$(a)\,\,\,\,\,\,\,\,\,\,\,\,\,\,\,\,\,\,\,\,\,\,\,\,\,\,\,\,\,\,\,\,\,\,\,\,\,\,\,\,\,\,\,\,\,\,\,\,\,\,\,\,\,\,\,\,\,\,\,\,\,\,\,\,\,\,\,\,\,\,\,\,\,\,\,\,\,\,\,\,\,\,\,\,\,(b)\,\,\,\,\,\,\,\,\,\,\,\,\,\,\,\,\,\,\,\,\,\,\,\,\,\,\,\,\,\,\,\,\,\,\,\,\,\,\,\,\,\,\,\,\,\,\,\,\,\,\,\,\,\,\,\,\,\,\,\,\,\,\,\,\,\,\,\,\,\,\,\,\,\,\,\,\,\,\,\,\,\,\,\,\,(c)$$
	\includegraphics[width=0.23\textwidth,angle=270]{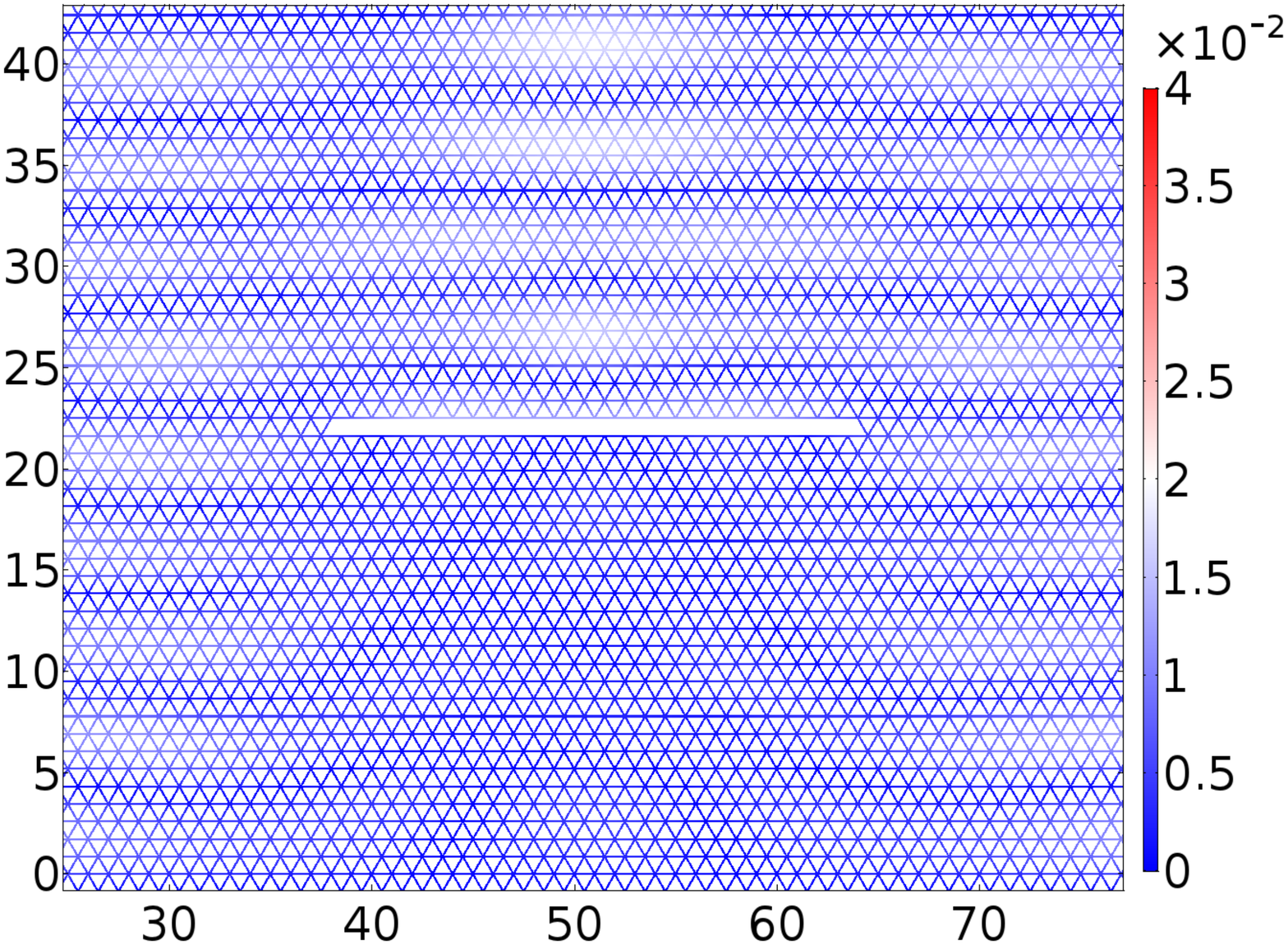}\hfill
	\includegraphics[width=0.23\textwidth,angle=270]{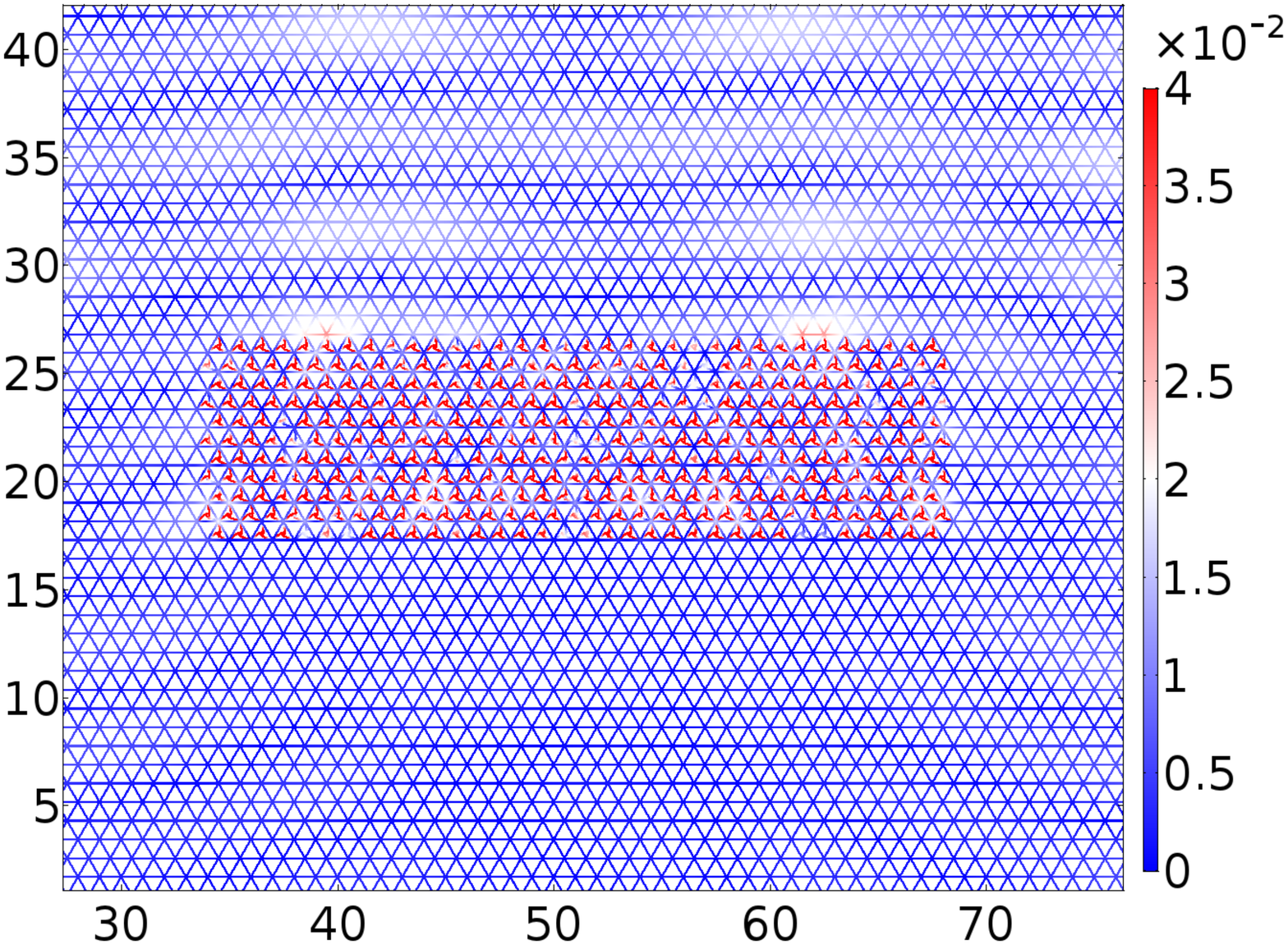}\hfill
	\includegraphics[width=0.23\textwidth,angle=270]{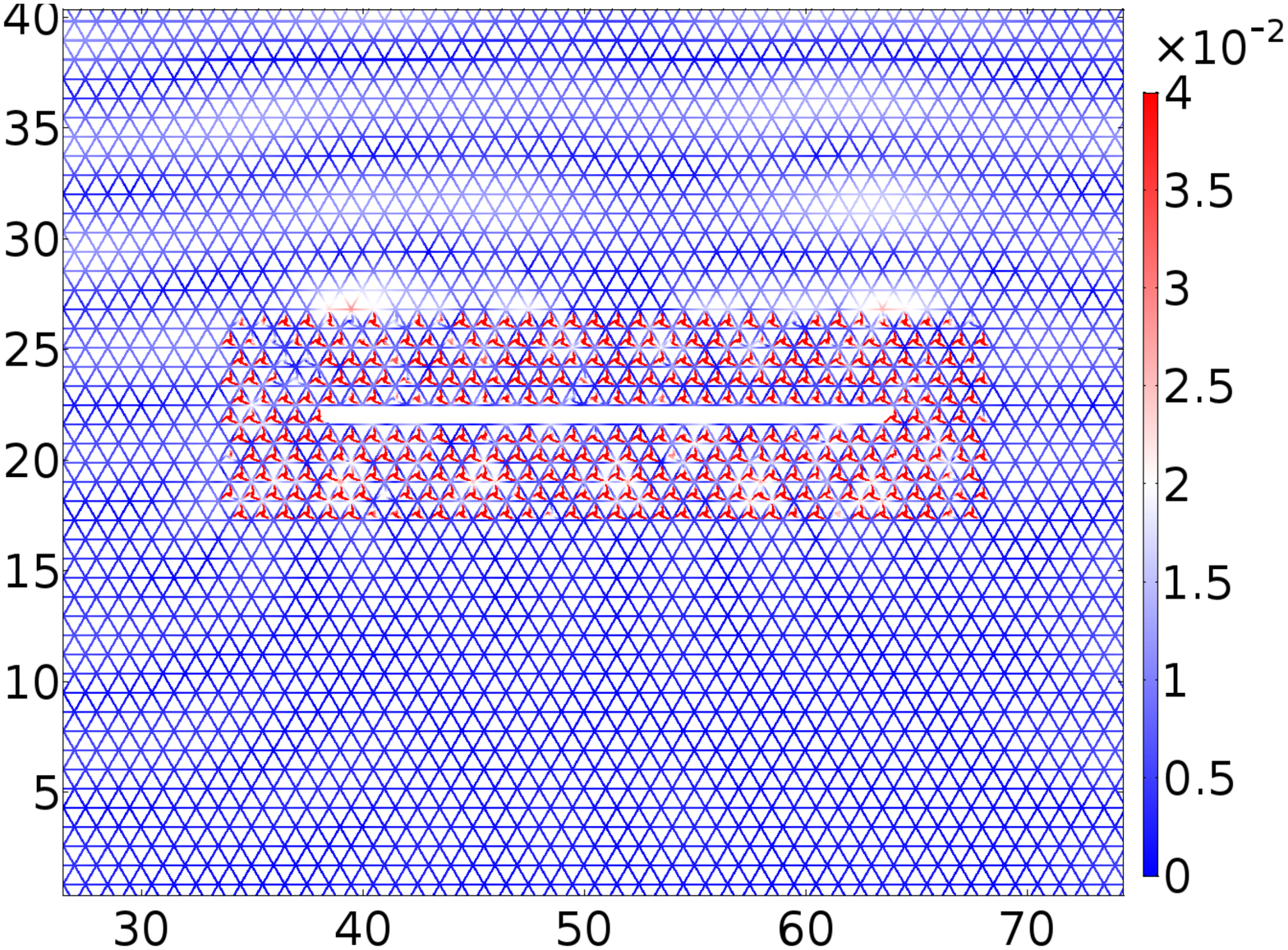}\\
	$$(d)\,\,\,\,\,\,\,\,\,\,\,\,\,\,\,\,\,\,\,\,\,\,\,\,\,\,\,\,\,\,\,\,\,\,\,\,\,\,\,\,\,\,\,\,\,\,\,\,\,\,\,\,\,\,\,\,\,\,\,\,\,\,\,\,\,\,\,\,\,\,\,\,\,\,\,\,\,\,\,\,\,\,\,\,\,(e)\,\,\,\,\,\,\,\,\,\,\,\,\,\,\,\,\,\,\,\,\,\,\,\,\,\,\,\,\,\,\,\,\,\,\,\,\,\,\,\,\,\,\,\,\,\,\,\,\,\,\,\,\,\,\,\,\,\,\,\,\,\,\,\,\,\,\,\,\,\,\,\,\,\,\,\,\,\,\,\,\,\,\,\,\,(f)$$
	\caption{\label{fig:EDGE} The harmonic response to a shear plane wave. Panels (a) and (d), (b) and (e), (c) and (f), comprise a crack, a cluster of resonators and a crack surrounded by a cluster of resonators, respectively. The angular frequency for panels (a), (b) and (c) is ${\omega}={\rm2.1~rad/s}$ corresponding to the lower edge of the band gap of Fig. \ref{fig:disp_DC}(a). The shear wave' s angular frequency for panels (d), (e) and (f) is ${\omega}={\rm2.7\,rad/s}$ corresponding to the upper edge of the band gap of Fig. \ref{fig:disp_DC}(a).}
\end{figure}
\subsection{Band gap regime \label{subsec:bg}}
In Figs \ref{fig:PW_T_Gap}, a shear plane wave coming from above impinges  at normal incidence on a cluster of resonators (panel (a)) and on clusters of resonators containing a cracks (panels (b),(c) and (d)). The frequency of the excitation is ${\omega}=2.4~{\rm rad/s}$ corresponding to the band gap of Fig. \ref{fig:disp_DC}(a). It is remarked that the coating is not penetrated by the incident wave. In particular, panel (b) shows that the structured cluster acts as a protective layer for the crack, as one would expect from the analysis of the dispersion diagram for Bloch waves. In Figs \ref{fig:PW_T_Gap}(c) and \ref{fig:PW_T_Gap}(d) we introduce a defect consisting of a missing line of resonators along the extension of the crack.  In Fig. \ref{fig:PW_T_Gap}(c) the tilting angle is homogeneous,  whereas in panel \ref{fig:PW_T_Gap}(d)  the resonators are tilted in opposite directions above and below the line defect. The stiffness of the links of the line defects \color{black} is the same as of the exterior triangular lattice. Figs \ref{fig:PW_T_Gap}(c) and  \ref{fig:PW_T_Gap}(d) show a displacement enhancement  at the perimeter of the cluster, however away from the crack tip. Fig. \ref{fig:PW_T_Gap}(e) highlights  an edge wave travelling along the boundary of the cluster. \color{black}

In Figs \ref{fig:EDGE}(a),  \ref{fig:EDGE}(b) and \ref{fig:EDGE}(c), the angular frequency ${\omega}=2.1~{\rm rad/s}$ of the plane wave corresponds to the lower edge of the band gap  of Fig. \ref{fig:disp_DC}(a).  In Figs \ref{fig:EDGE}(d),  \ref{fig:EDGE}(e) and \ref{fig:EDGE}(f), the frequency  ${\omega}=2.7~{\rm rad/s}$ corresponds to the upper edge of the band gap. For the lower edge frequency, although the cluster is partially protective (see panel (b)), the introduction of the one-dimensional defect increases the stress concentration around the crack (panels (c)), compared to the uncoated configuration (panels (a)). \color{black} A similar effect is reported for the upper edge of the band gap in Fig. \ref{fig:EDGE}(e) and Fig. \ref{fig:EDGE}(f). In the vicinity of the band gap edges, the coating of resonators enhances the displacement field around the crack, increasing the chances for the crack to propagate. 

\begin{figure}[h!]
	\centering
	\includegraphics[width=0.5\textwidth]{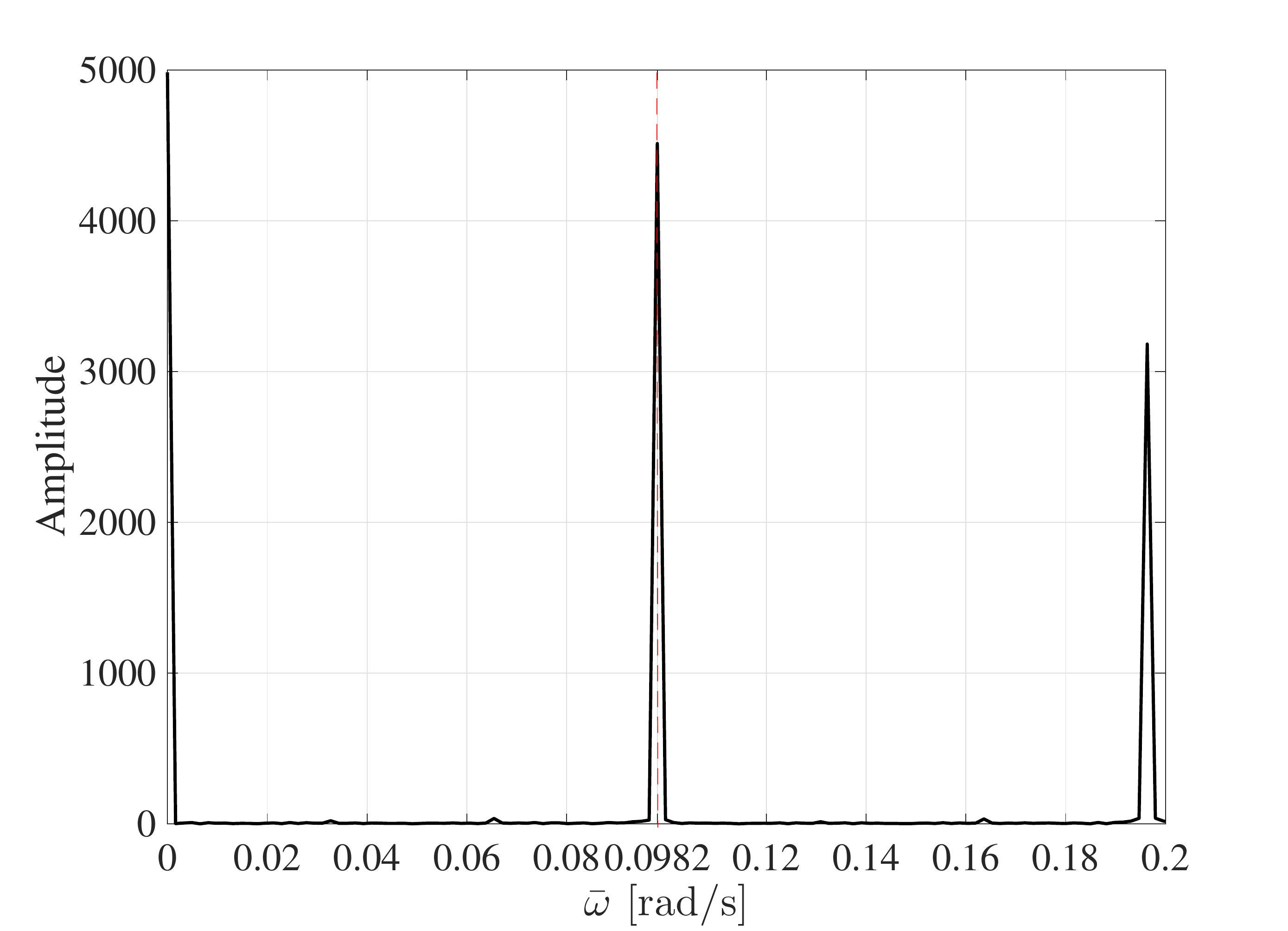}
	\caption{\label{fig:alessio_FT}  
		The Fast Fourier Transform of the input pulsating load has identified a countable number of spikes at different frequencies. Two spikes in the low frequency regime are shown here. }
\end{figure}
\begin{figure}[h!]
	\centering
	\includegraphics[width=0.33\textwidth]{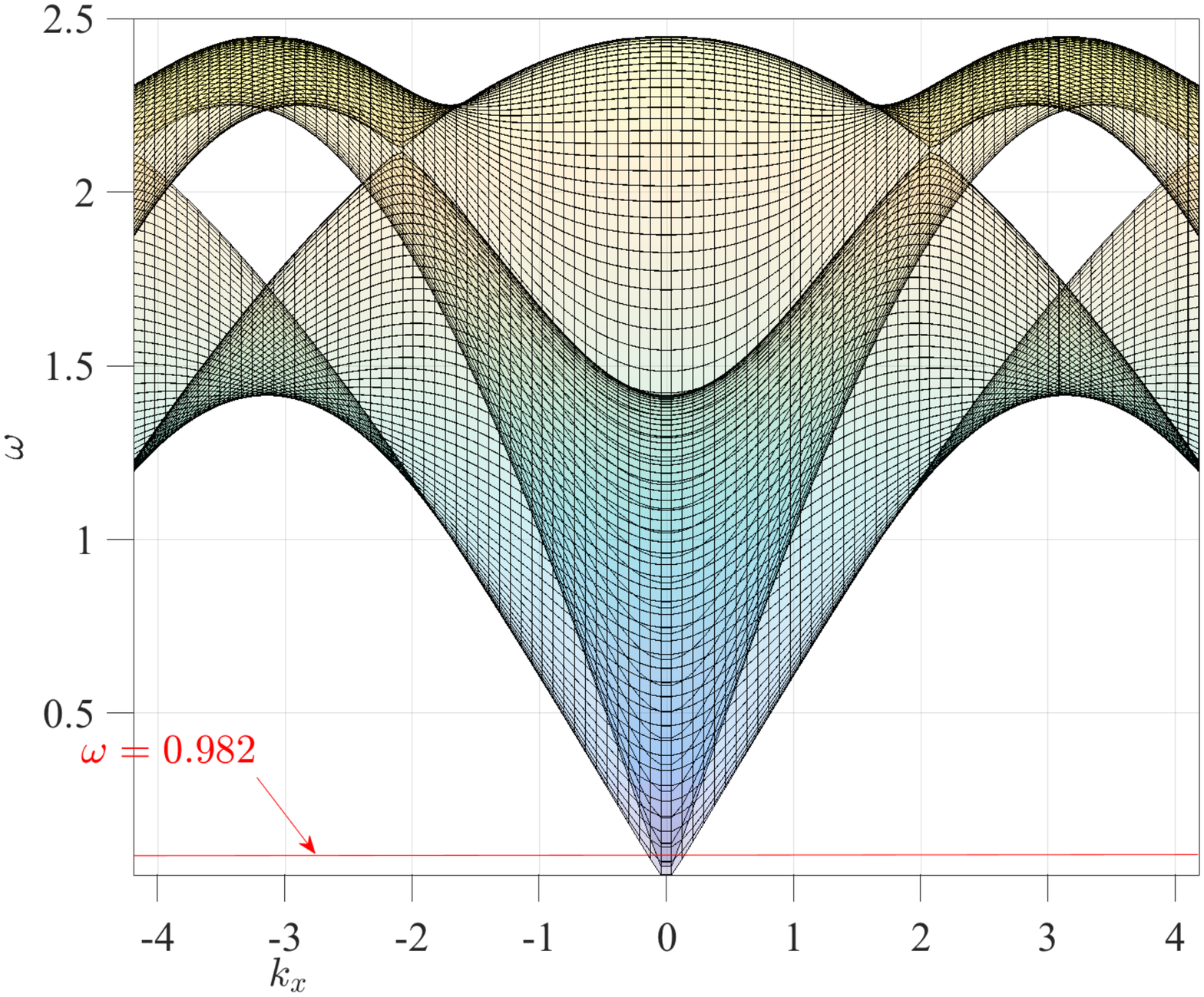}\hfill
	\includegraphics[width=0.33\textwidth]{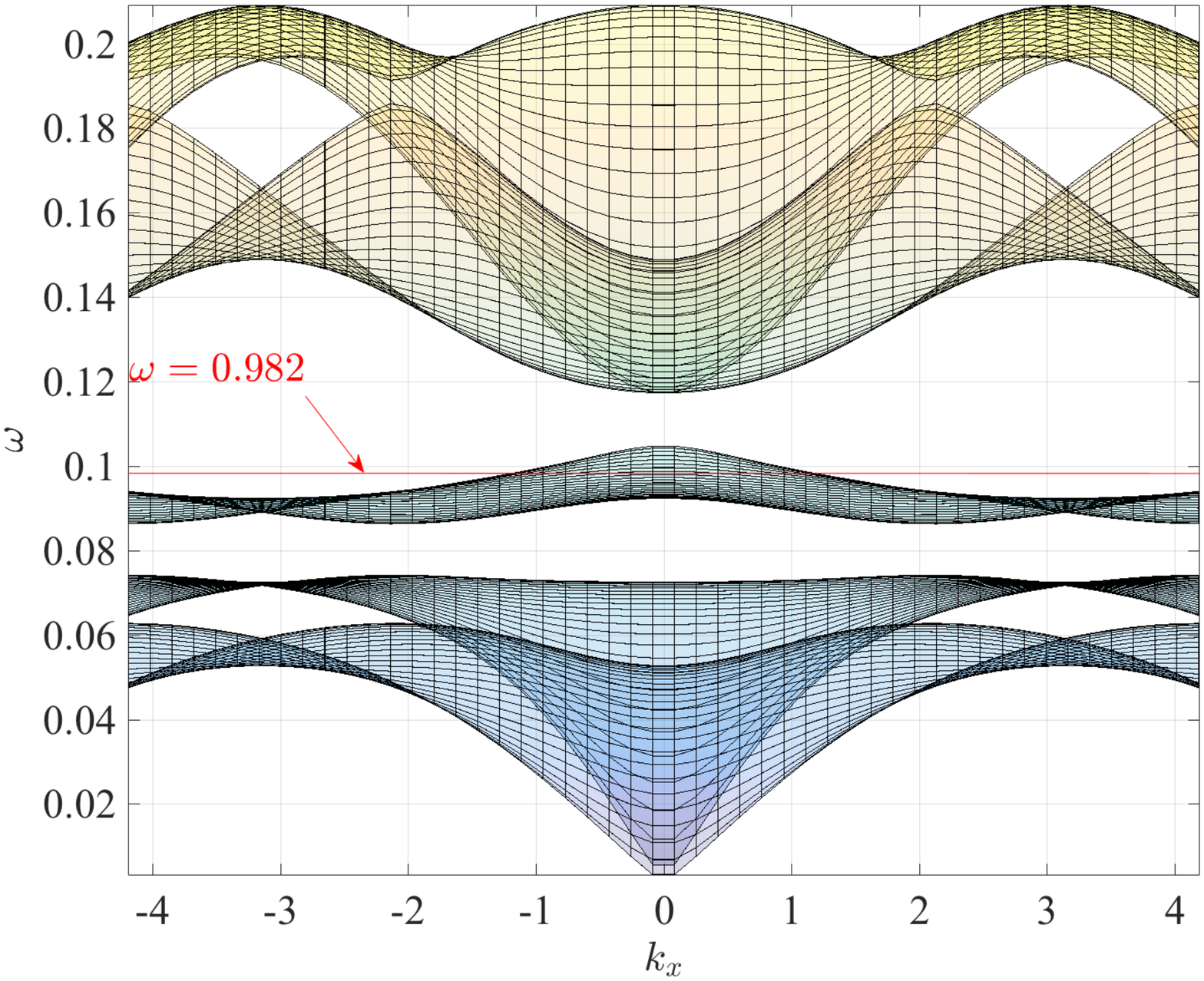}\hfill
	\includegraphics[width=0.33\textwidth]{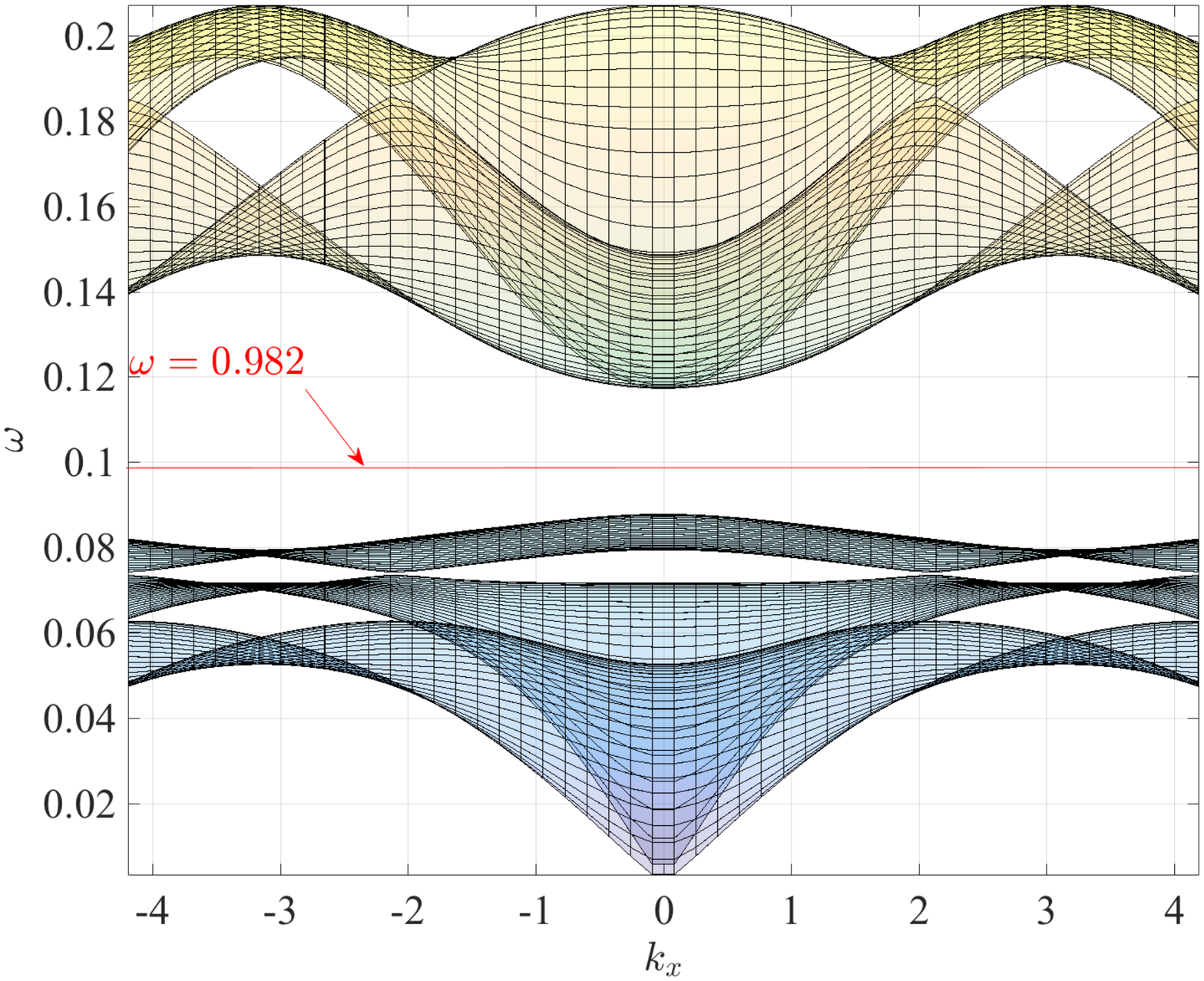}
	$\,\,\,\,\,\,\,\,\,\,\,\,\,\,\,\,\,\,\,\,\,\,\,\,\,\,\,\,\,\,\,\,\,\,\,\,\,\,(a)\,\,\,\,\,\,\,\,\,\,\,\,\,\,\,\,\,\,\,\,\,\,\,\,\,\,\,\,\,\,\,\,\,\,\,\,\,\,\,\,\,\,\,\,\,\,\,\,\,\,\,\,\,\,\,\,\,\,\,\,\,\,\,\,\,\,\,\,\,\,\,\,\,\,(b):~78~{\rm deg}\,\,\,\,\,\,\,\,\,\,\,\,\,\,\,\,\,\,\,\,\,\,\,\,\,\,\,\,\,\,\,\,\,\,\,\,\,\,\,\,\,\,\,\,\,\,\,\,\,\,\,\,\,\,\,\,\,\,\,\,(c):~47~{\rm deg}$
	\caption{\label{fig:disp_Alessio} Dispersion surfaces for a TL (panel (a)) and for two TLRs (panels (b) and (c)). Panel (a) has been obtained using the parameters listed in the third row of Table \ref{tab:parameters_alessio}. Panels (b) and (c) correspond to the parameters listed in the second and first rows of   Table \ref{tab:parameters_alessio}, respectively.}
\end{figure}

\section{Edge crack subjected to a transient thermal load \label{sec:crack_trans}}
The governing equations, loading configuration and the fracture criterion are  the same as in the earlier computations for the thermoelastic crack advancing through a homogeneous triangular lattice \cite{trevisan_2016}. Here, a geometrically chiral coating surrounding the crack is introduced into the model.  
 An elastic wave is generated as a result of a rapid variation of the boundary temperature. The fracture criterion is based on a normalised threshold elongation $\epsilon=\Delta L / L$. The crack advances when the ligament at the crack tip reaches the critical threshold elongation.  The loading configuration is made of square pulses applied to the left edge of the computational domain. The  period of the load is $\theta=4 \tau$, where $\tau=16~{\rm s}$ is the duration of a single pulse.  The radian frequency of the pulse is $\omega_s=2\pi/\theta=0.0982~{\rm rad/s}$, where the subscript $s$ stands for  ``striping". The duration of the pulse is $60~\theta$. Fig.  \ref{fig:alessio_FT} shows the Fourier spectrum of the temperature loading. We observe that the spectrum is dominated by spikes occurring at multiples of $\omega_s$. We limited the plot to $\bar{\omega} \in[0,2.1 \omega_s] $, where the most pronounced spikes of the spectrum appear. 
\begin{table}[t!]
	\centering
	\begin{tabular}{@{}llllllllr@{}} \toprule
		&$c_\ell~[{\rm N/m}]$ & $m~[{\rm Kg}]$ & $L~[{\rm m}]$ & $\ell~[{\rm m}]$ & $c_{\ell o}~[{\rm N/m}]$ &$m_o~[{\rm Kg}]$& $\vartheta_{0}~[{\rm deg}]$ & $\alpha[{\rm C}^{-1}]$\\ \midrule
		${\rm set~1 ,~TLR}$&1            &            181.82      & 1    &0.21 &  1    &        90.91     & 47& $10^{-3}$\\
		${\rm set~2 ,~TLR}$&1            &           181.82      & 1    &0.21 &  1    &        90.91      & 78 & $10^{-3}$\\
		${\rm TL}$&1            &                  1&  1   & &     &   && $10^{-3}$ \\\bottomrule
	\end{tabular}\caption{\label{tab:parameters_alessio}Thermoelastic parameters for the ambient triangular lattice (third row) and for triangular lattices with resonators (first and second rows). The parameter $\alpha=({\rm d}L/{\rm d}T)/L$ is the longitudinal coefficient of thermal expansion which applies to the triangular lattice links only. The remaining links are such that $\alpha=0$.}
\end{table}

Tilted resonators are added as four layers (two above and two below the crack). The trusses which link the resonators to the TL's nodal points are thermally insulating.  The mass of the unit cell containing a resonator is not equal to the mass of the exterior triangular lattice nodal points. In Table \ref{tab:parameters_alessio}, we list the thermoelastic parameters used in the  transient non-linear simulations. The dispersion diagrams corresponding to the periodic lattices are represented in Fig. \ref{fig:disp_Alessio}. Fig. \ref{fig:disp_Alessio}(a) represents the dispersion surfaces for the triangular lattice outside the  cracked strip. Figs  \ref{fig:disp_Alessio}(b) and \ref{fig:disp_Alessio}(c) show the dispersion diagrams for two  triangular lattices with resonators which differ from each other by the tilting angle ($47~{\rm deg}$ and $78~{\rm deg}$, respectively). The structured lattices are deliberately designed in such a way that $\omega_s$ lies in the passband for Fig. \ref{fig:disp_Alessio}(b) and in the stop band for Fig. \ref{fig:disp_Alessio}(c), as highlighted by the horizontal red lines.

\begin{figure}[t!]
	\centering
	\includegraphics[width=0.5\textwidth]{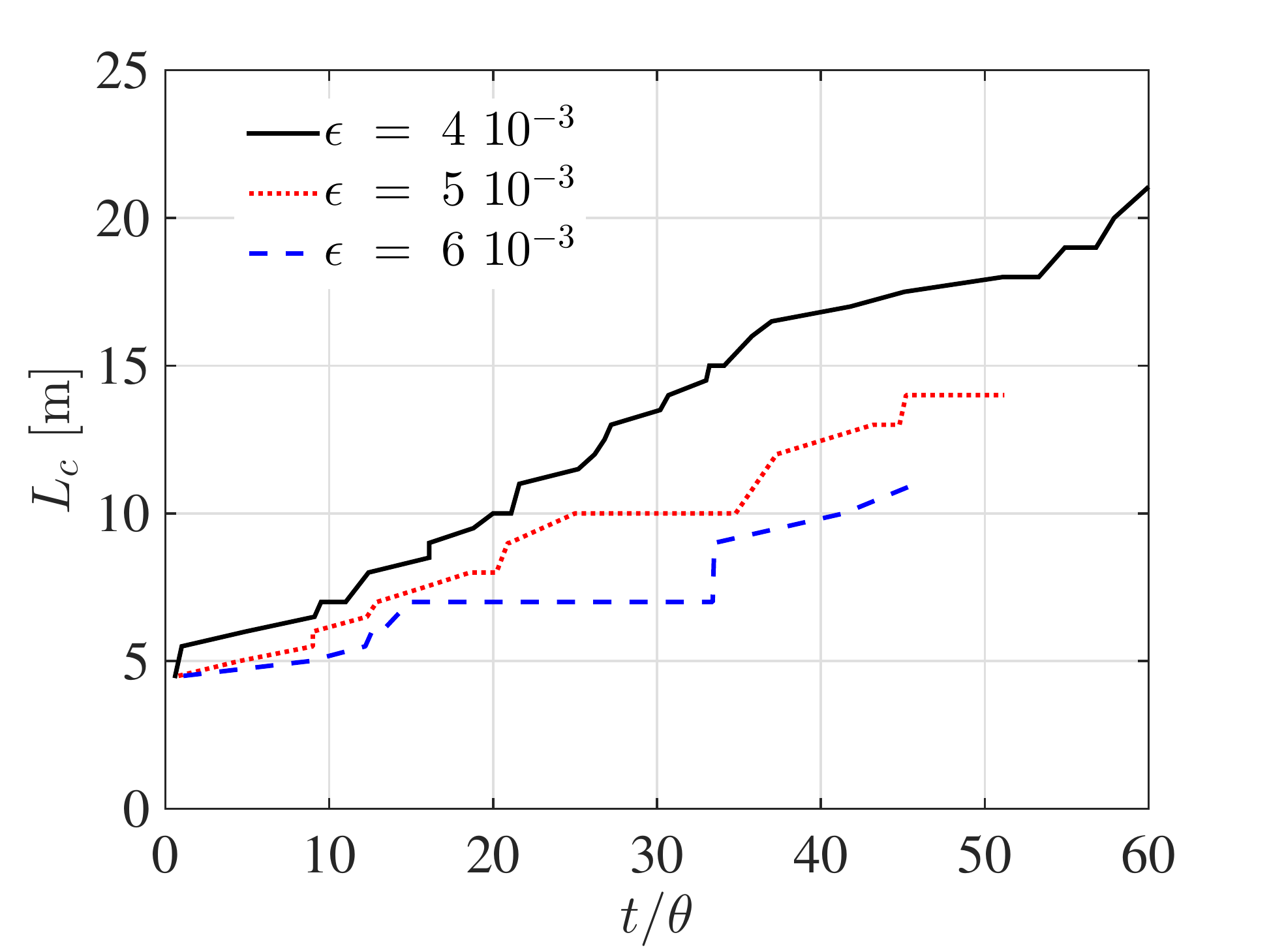}\hfill
	\includegraphics[width=0.5\textwidth]{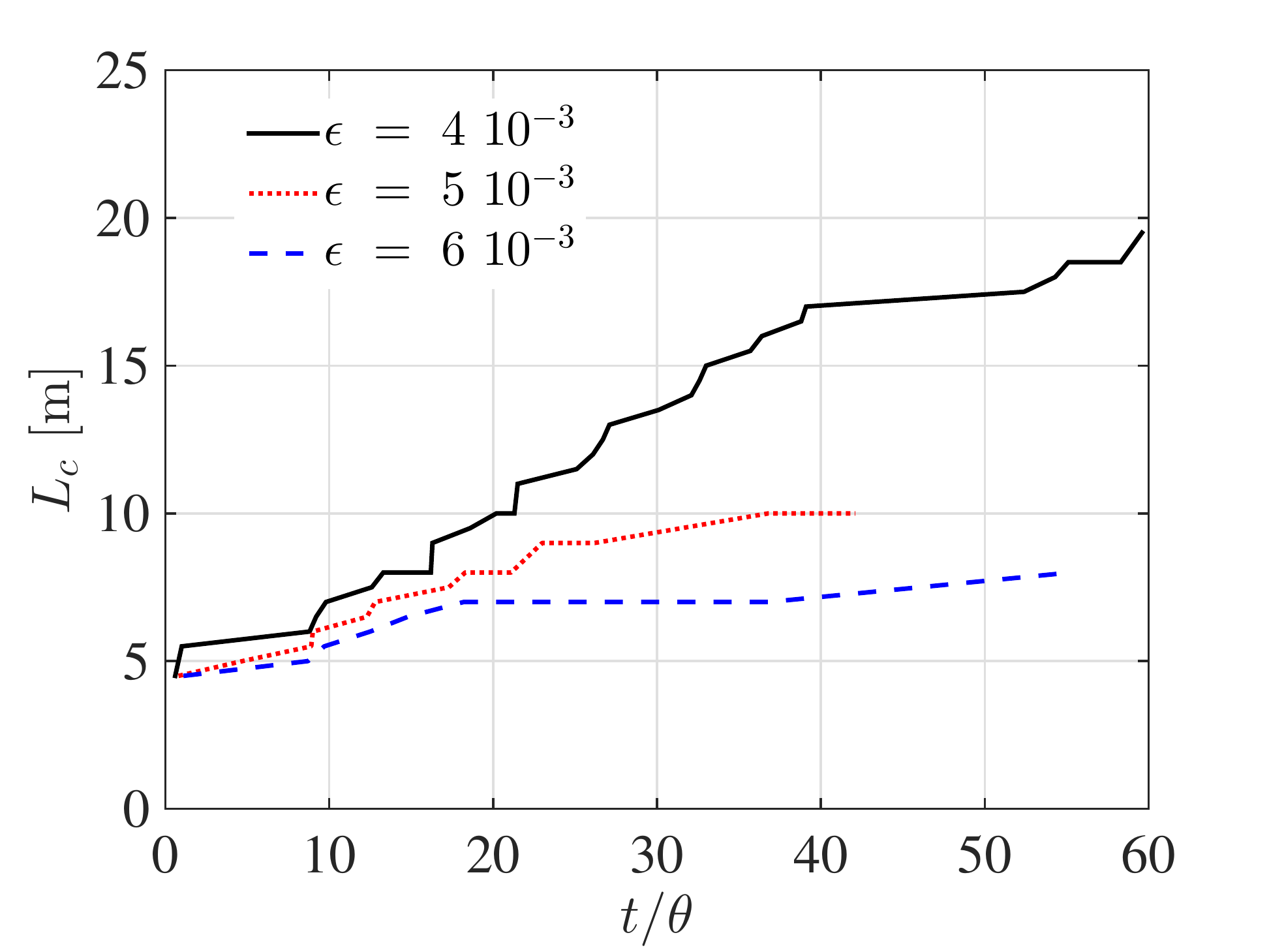}
$(a):\vartheta_0=47~{\rm deg}\,\,\,\,\,\,\,\,\,\,\,\,\,\,\,\,\,\,\,\,\,\,\,\,\,\,\,\,\,\,\,\,\,\,\,\,\,\,\,\,\,\,\,\,\,\,\,\,\,\,\,\,\,\,\,\,\,\,\,\,\,\,\,\,\,\,\,\,\,\,\,\,\,\,\,\,\,\,\,\,\,\,\,\,\,\,\,\,\,\,(b):\vartheta_0=78~{\rm deg}$
\caption{\label{fig:crack_speed} Crack length $L_c$ as a function of  time for two configurations corresponding to two tilting angles. The hosting triangular lattice and the parameters for the two lattices with resonators are reported in Table  \ref{tab:parameters_alessio}. }
\end{figure}

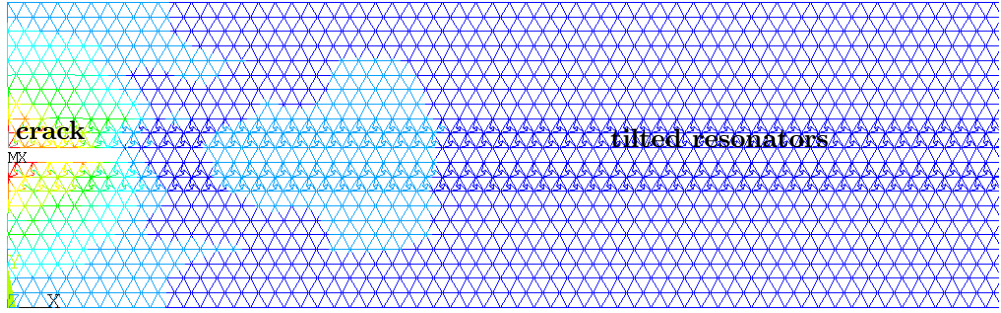
\begin{figure}[h!]
	\centering
	\includegraphics[width=0.8\textwidth]{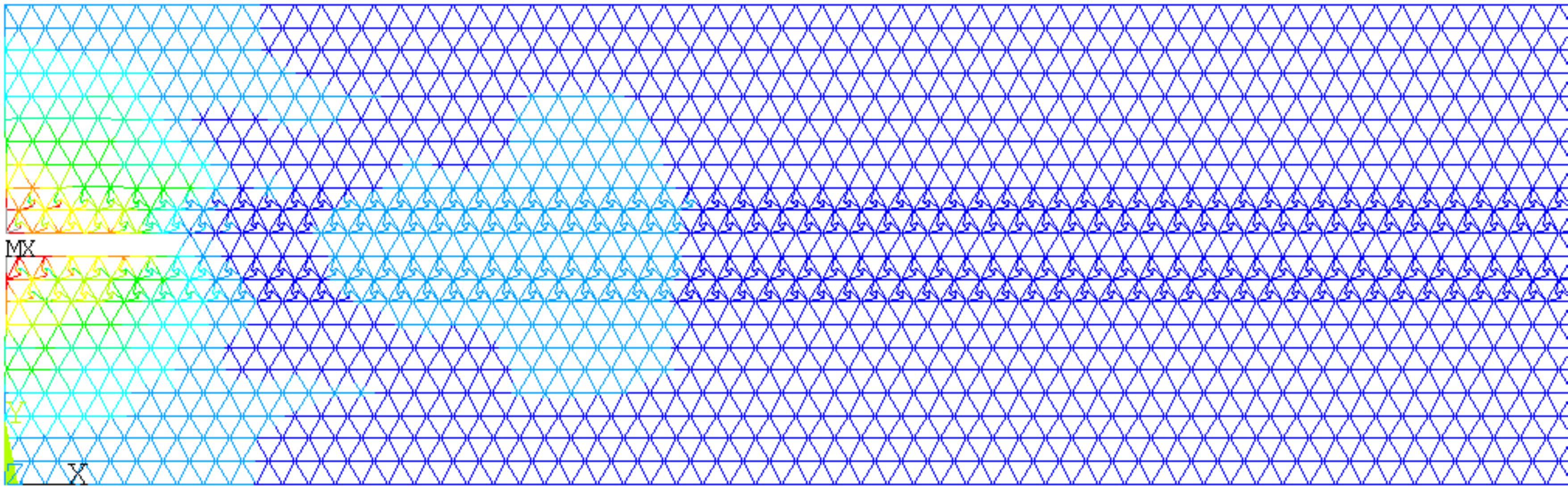}
	\begin{picture}(0,0)(0,0)
	\put(-375,65) {\mbox{\bf crack}}
	\put(-150,62) {\mbox{\bf tilted resonators}}
	\end{picture}
	\caption{\label{fig:crack_displ} Instantaneous modulus of the displacement for the time step $t/\theta\approx10$  in the transient simulation  represented in Fig. \ref{fig:crack_speed}(b) for the  elongation threshold $\epsilon=5\cdot10^{-3}$ (red dotted line). The crack tip ``emits" elastic waves which propagate along the coating.}
\end{figure}

From the transient solution of the thermoelastic problem described above, we extracted the crack length $L_c$ at several times. The results are represented in Fig. \ref{fig:crack_speed} for different normalised elongation thresholds $\epsilon$. Panel (a) corresponds to the lower tilting angle and panel (b) to the higher one. At the same elongation thresholds, the average crack speeds in panel (a) are slightly higher than those in panel (b). 
We provide a qualitative interpretation of this phenomenon as follows.
The thermal shocks trigger elastic waves whose amplitudes \emph{vs} frequency at the left edge of the computaional window, differ from Fig. \ref{fig:alessio_FT} by a multiplicative constant. When $\omega_s$ is in the passband, \emph{i.e.} when $\vartheta_0=78~{\rm deg}$, elastic waves can propagate along the strip of tilted resonators (see Fig. \ref{fig:disp_Alessio} ), resulting in a reduction of strain concentration at the crack tip compared to the $\vartheta_0=47~{\rm deg}$ configuration.  Equivalently, the strip of resonators acts as a  structured waveguide which channels the energy away from the crack tip, as illustrated in Fig.  \ref{fig:crack_displ}.  On the contrary, when $\vartheta_0=47~{\rm deg}$, the waveguide action is being suppressed, which leads to the field localisation around the crack and hence the stronger advance of the fracture through the lattice. 

\section{Concluding remarks \label{sec:conclusion}} 

We have identified several important applications of a novel geometrically chiral micro-structure in the design of advanced materials, used as filters/polarisers of elastic waves.  

A transient advance of a crack, whose instantaneous snapshot is given in Fig. \ref{fig:crack_displ}, has been studied in a micro-structured layer where tilted resonators in the lattice are present. The analysis of the transient crack advance illustrated by  Figs \ref{fig:crack_speed}(a) and \ref{fig:crack_speed}(b) is linked to the tunable dispersion properties of the lattices (see Fig. \ref{fig:disp_Alessio}) and to the guiding features of the structured coating around the crack, as shown in  Fig. \ref{fig:crack_displ}.    

The Dirac-like dynamic regime deserves the special mention. It has been achieved and studied here in relation to the wave-guiding and wave-defect interaction problems. Asymmetries in the scattered elastic field have been identified for waves at the Dirac-like frequency. This in turn empowers further studies in the context of asymmetric crack initiation mechanisms (see Figs \ref{fig:DC_T} and \ref{fig:DC_long}). 

Shielding of a defect from an incident elastic shear wave have been achieved in the regimes, which correspond to the complete band-gap of the triangular lattice with resonators. In addition to the usual low penetration of  external waves within the protecting coating, we emphasise  that edge waves occur around the perimeter of the coating  in our model (see Fig. \ref{fig:PW_T_Gap}). This is a ``finger-print" of the lattice's geometric chirality, and cannot be achieved by  the straightforward adjustment of the triangular lattice parameters, \emph{e.g.} by introducing a contrast in the inertia or stiffness. 
\section*{Acknowledgements}
D.T. gratefully acknowledges the People Programme (Marie Curie Actions) of the European Union's Seventh Framework Programme FP7/2007-2013/ under REA grant agreement number PITN-GA-2013- 606878. The paper was completed while D.T. was in a work secondment at Enginsoft (Italy), whose stimulating and welcoming environment is gratefully acknowledged. A.B.M. and N.V.M. acknowledge the financial support of the EPSRC through programme grant EP/L024926/1. The paper was completed while A.B.M. was visiting the University of Trento;  the support from the ERC Advanced Grant ‘Instabilities and nonlocal multiscale modelling of materials’ FP7-PEOPLE-IDEAS-ERC-2013-AdG is gratefully acknowledged.
\bibliographystyle{unsrt}
\bibliography{bibliography}{}
\bibliographystyle{unsrt}
\end{document}